 \documentclass[twocolumn,trackchanges]{aastex631}

\shorttitle{Identifying Variable mid-L dwarfs}
\shortauthors{Oliveros-Gómez et al.}
\graphicspath{{./}{figures/}}
\usepackage{booktabs}
\usepackage{rotating}
\usepackage{verbatim} 
\usepackage[normalem]{ulem}
\usepackage{booktabs, longtable}
\usepackage{savesym}
\savesymbol{tablenum}
\usepackage{siunitx}
\restoresymbol{SIX}{tablenum}
\usepackage{nicefrac}
\usepackage[flushleft]{threeparttable}

\begin{document}

\title{An Informed and Systematic Method to Identify Variable mid-L dwarfs}

\author[0000-0001-5254-6740]{Natalia Oliveros-Gomez}
\affiliation{Departamento de Astronomía, Universidad de Guanajuato. Callej\'on de Jalisco, S/N, 36023, Guanajuato, GTO, M\'exico}
\affiliation{William H. Miller III Department of Physics and Astronomy, Johns Hopkins University, Baltimore, MD 21218, USA}

\author[0000-0003-0192-6887]{Elena Manjavacas}
\affiliation{AURA for the European Space Agency (ESA), ESA Office, Space Telescope Science Institute, 3700 San Martin Drive, Baltimore, MD, 21218 USA}
\affiliation{William H. Miller III Department of Physics and Astronomy, Johns Hopkins University, Baltimore, MD 21218, USA}

\author[0000-0001-8170-7072]{Daniella C. Bardalez Gagliuffi}
\affiliation{Department of Physics \& Astronomy, Amherst College, 25 East Drive, Amherst, MA 01003, USA}
\affiliation{Department of Astrophysics, American Museum of Natural History 
200 Central Park West, New York, NY 10024, USA}

\author[0000-0001-7356-6652]{Theodora Karalidi}
\affiliation{Department of Physics, University of Central Florida, 4000 Central Florida Blvd., Orlando, FL 32816, USA}

\author[0000-0003-0489-1528]{Johanna M. Vos}
\affiliation{School of Physics, Trinity College Dublin, The University of Dublin, Dublin 2, Ireland}
\affiliation{Department of Astrophysics, American Museum of Natural History, New York, NY 10024, USA}

\author[0000-0001-6251-0573]{Jacqueline K. Faherty}
\affiliation{Department of Astrophysics, American Museum of Natural History 
200 Central Park West, New York, NY 10024, USA}



\begin{abstract}

Most brown dwarfs show some level of photometric or spectral variability. However, finding the most variable dwarfs more suited for a thorough variability monitoring campaign remained a challenge until a few years ago with the design of spectral indices to find the most likely L and T dwarfs using their near-infrared single-epoch spectrum. In this work, we designed and tested near-infrared spectral indices to pre-select the most likely variable L4-L8 dwarfs, complementing the indices presented by \cite{ashraf2022disentangling} and \cite{oliveros2022informed}. We used time-resolved near-infrared \textit{Hubble Space Telescope} \textit{Wide Field Camera 3} spectra of a L6.0 dwarf, LP~261--75b, to design our novel spectral indices. We tested these spectral indices on 75 L4.0-L8.0 near-infrared SpeX/IRTF  spectra, providing 27 new variable candidates. Our indices have a recovery rate of $\sim$80\%, and a false negative rate of $\sim$25\%. All the known non-variable brown dwarfs were found to be non-variable by our indices. We estimated the variability fraction of our sample to be $\mathrm{51^{+4}_{-38}}$\%, which agrees with the variability fractions provided by \cite{buenzli2014brown}, \cite{radigan2014strong} and \cite{metchev2015weather} for L4--L8 dwarfs. These spectral indices may support in the future the selection of the most likely variable directly-imaged exoplanets for studies with the \textit{James Webb Space Telescope} and as well as the 30-m telescopes.

\end{abstract}

\keywords{Variability --- Brown dwarfs --- L dwarfs --- Spectral indices }


\section{Introduction} \label{sec:intro}

Since the discovery of brown dwarfs almost three decades ago \citep{Rebolo1995, Oppenheimer1995}, multiple surveys have searched for photometric and spectro-photometric variability of brown dwarfs (e.g. \citealt{Bailer_Jones1999, Clarke2008, Artigau2009, radigan2014strong, metchev2015weather}, among others) with ground-based and space-based telescopes, concluding that, up to some extent, the great majority of brown dwarfs show some level of photometric or spectro-photometric variability. Although other scenarios are also possible \citep{tremblin2015fingering, tremblin2016cloudless}, the most likely cause of this variability is the presence of heterogeneous clouds in the atmospheres of brown dwarfs at different pressure levels (e.g. \citealt{apai2013hst}).  

The variability amplitudes measured for all the brown dwarfs that were monitored to date vary from object to object (see \citealt{vos2020spitzer} for a compilation), although L/T transition brown dwarfs and low surface gravity brown dwarfs usually show higher variability amplitudes in general \citep{radigan2014strong, metchev2015weather, vos2018variability}. \cite{Vos2017} also concluded that the variability amplitude depends on the viewing angle at which one specific object is observed. Brown dwarfs with equator on viewing angles show in general higher variability amplitudes than those observed pole on \citep{Vos2017}. In addition, General Circulation models (GCMs) of brown dwarfs \citep{Tan_Showman2021} also agree that equator-on brown dwarfs should show high variability amplitudes. Finally, recent work by \cite{Suarez2023} further confirmed the latitudinal dependence of dust cloud opacity in ultracool L dwarfs, supporting the idea that the equatorial latitudes are cloudier than polar latitudes of brown dwarfs, which might explain the higher variability amplitudes observed for equator-on brown dwarfs \citep{Vos2017}.
It is also important to note that brown dwarf light curves and variability amplitudes were found to evolve with time, changing their shape and variability amplitudes even only within a few rotational periods \citep{apai2017zones, Apai2021}, indicating the existence of evolving cloud patterns, like high-speed jets, zonal circulation and vortexes, in the atmospheres of brown dwarfs.

Most field mid-L dwarfs usually show some level of variability with relatively small amplitudes between 0.2\% and 1.5\% ($\mathrm{80^{+20}_{-27}}$\%, according to \citealt{metchev2015weather}). Field mid-L dwarfs have been found to have higher variability amplitudes, but none have been found to have extremely variable light curves ($>$10\%) as for their younger counterparts ($<$100~Myr). However, brown dwarfs with a relatively high variability amplitude ($>$1\%) are the best for more in-depth characterization which would allow us, for example, to do a robust mapping of the surface of the object \citep{karalidi2015aeolus}. Nonetheless, obtaining light curves for brown dwarfs is a resource-intensive task, since in general at least a few hours of telescope monitoring are required per object to be able to obtain a useful light curve. Thus, selecting the most likely variable brown dwarfs is key to optimally employing resources. Until \cite{ashraf2022disentangling} and \cite{oliveros2022informed} designed a systematic method based on spectral indices to pre-select the most likely variable L/T and T dwarfs, the search for photometric variability of brown dwarfs was completely uninformed, making variability searches in brown dwarfs a risky endeavor.

In this paper, we designed a novel set of spectral indices to find variable candidates with mid-L (L4-L8) spectral types employing a single-epoch near-infrared spectrum. We employed time-resolved near-infrared spectra of 2MASSW~J09510549+3558021B (LP267-75B) obtained with the \textit{Wide Field Camera 3} instrument (WFC3), onboard the \textit{Hubble Space Telescope} (HST) \citep{manjavacas2017cloud} for designing our spectral indices. This work complements the spectral indices presented in \cite{ashraf2022disentangling} and \cite{oliveros2022informed} to allow the pre-selection of the most likely variable mid-L dwarfs. 

This paper is structured as follows: In Section \ref{sec:observations} we provide the details of the \textit{HST/WFC3} observations for LP~261-75b. In Section \ref{sec:index-method}, we explain the design of the near-infrared spectral indices for detecting variability in L type dwarfs. In Section \ref{Sec:Candidate-variables}, we compare our results with previous studies. In Section \ref{Sec:discussion}, we discuss the variability fraction obtained for our sample and the physics behind the spectral indices. Finally, in Section \ref{Sec:conclusions}, we summarize our conclusions.

\section{LP 261--75B}

LP~267-75b is an L6 brown dwarf in optical \citep{kirkpatrick200067}. \cite{reid2006lp}, companion to an M4.5V star, with an estimated age of 100–200 Myr based on the primary star's coronal activity and separated by $\sim$450~AU \citep{reid2006lp}, and a mass around $M\sim 0.02^{+0.01}_{-0.005} M_{\odot}$ using evolutionary models. \citep{bowler2013planets} found an effective temperature of $\mathrm{T_{eff}} =1500$~K, and a surface gravity of log(g) = 4.5~dex (intermediate to field gravity, by fitting the Spectral Energy Distribution (SED) of the object using the BT-Settl atmospheric models. 

LP~267-75b is a variable dwarf with a period of 4.78 $\pm$ 0.95 hr \citep{manjavacas2017cloud}, with a minimum rotational modulation amplitude of approximately $2.41\% \pm 0.14\%$ on its white light curve. The amplitude variations inside the H$_2$O band (1.35–1.43 $\mu$m) and outside the H$_2$O band are consistent with the overall variability ($2.80\% \pm 1.02\%$).

\section{OBSERVATIONS}\label{sec:observations}

The data used for our analysis were obtained on  December 21, 2016 (UTC) with the infrared channel of the \textit{Wide Field Camera 3} (WFC3) on board the \textit{Hubble Space Telescope} (HST) and its G141 grism, in Cycle 23 of the HST (P.I. D. Apai, GO14241), all of the data can be found in MAST: \dataset[10.17909/jkbe-zn98]{http://dx.doi.org/10.17909/jkbe-zn98}. The wavelength spectral coverage of the HST/WFC3 spectra range between 1.05 and 1.69 $\mu$m. 

The observations were performed using a 256 x 256 pixels subarray of the WFC3 infrared channel, with a plate scale of 0.13 arcsec/pixel. In addition, a direct image was taken at each orbit in the F132N filter to obtain an accurate wavelength reference. The spectra were acquired over six consecutive HST orbits, spanning 8.49~hr, obtaining a total of 66 spectra. In each orbit were obtained a total of 11 spectra with a resolving power of 130 at 1.4 $\mu$m. The reduced spectra used for this work were published in \cite{manjavacas2017cloud}.

\section{Index Method}\label{sec:index-method}

\subsection{Designing Spectral indices}

\begin{figure}
    \centering
    \includegraphics[width=0.9\linewidth]{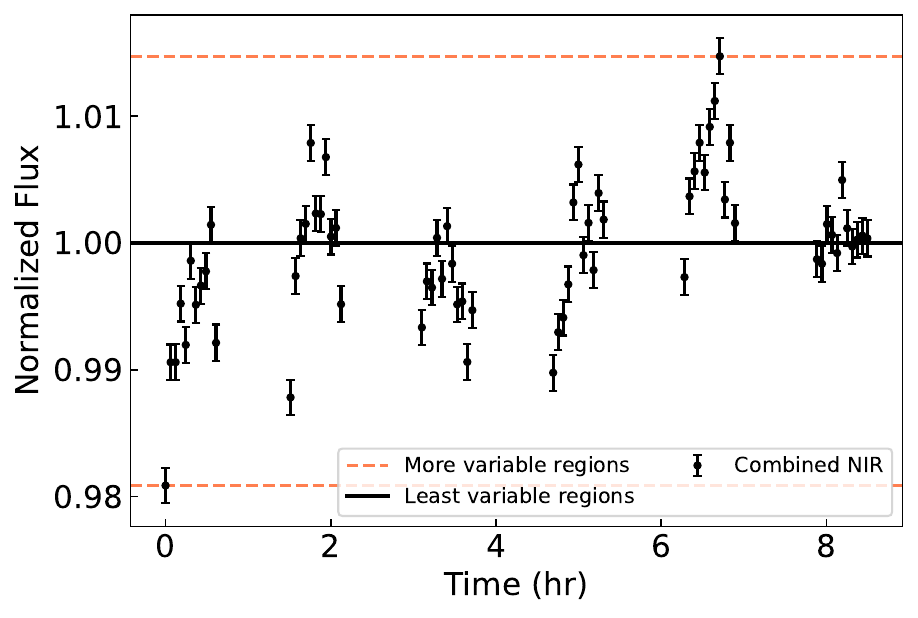}
    \caption{HST/WFC3 near-infrared light curve of LP~261-75b. Each data point corresponds to one spectrum of the object at a given phase of its rotational period. We mark with a black line the center of the light curve (normalized flux = 1), the spectra corresponding to the points of the light curve close to 1, are the spectra that show the flux minimum spectrum. The flux maximum spectra are those corresponding to the peaks of the light curve (orange dashed lines).}
    \label{fig:LC_lp261}
\end{figure}

To design our spectral indices, we follow the same procedure as for \cite{oliveros2022informed}. In this Section, we summarize the procedure. In Figure \ref{fig:LC_lp261} we show the near-infrared white light curve of LP~261-75b as published in \cite{manjavacas2017cloud}. The spectra with the highest variability are those at the extremes of the light curve and those with the lowest variability are those closest to unity. Those spectra will be used in our index method to spot the spectral characteristics that vary the most across the entire spectrum.  For this, we produce a template spectrum with the median combining all 66 spectra of the object. This spectrum will be equivalent to a higher signal-to-noise, single-epoch spectrum similar to those normally found in spectral libraries of brown dwarfs. We subtract each of the spectra in the light curve from this template to quantitatively find the differences of each spectrum in each wavelength range concerning the template.

\begin{figure}
    \centering
    \includegraphics[width=0.92\linewidth]{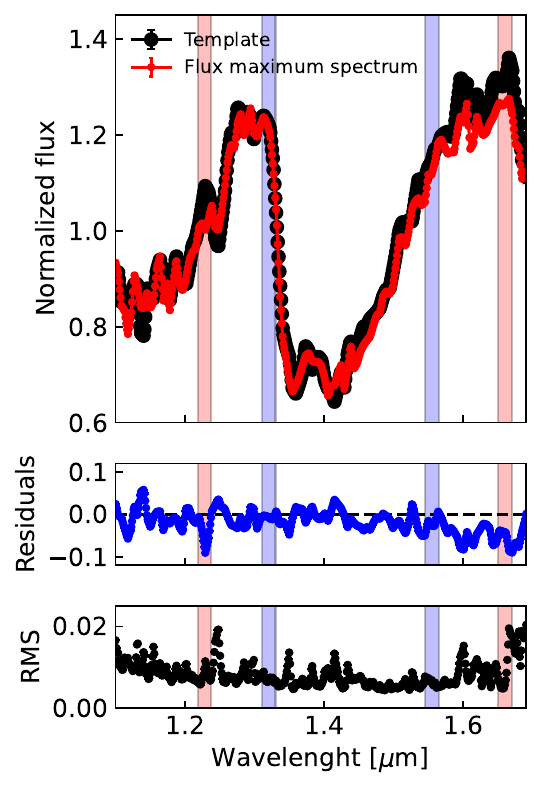}
    \caption{Top panel: \textit{template} of LP~261-75b after combining the median of the 66 HST/WFC3 spectra (black), and the flux maximum spectrum for LP~261-75b. Medium panel: residuals after subtracting the variable spectrum of LP~261-75b with the \textit{template} of LP~261-75b.  Lower panel: RMS of the timeseries (66 spectra) of LP261-75b against wavelength. We highlight in red the most variable region, and in blue the least variable regions of the spectrum according to the residuals shown.}
    \label{fig:residuals}
\end{figure}

In Figure \ref{fig:residuals}, upper panel, we show the flux maximum spectrum (in red) compared to the template (in black), and their respective residuals at the bottom panel. Windows of 0.03~$\mu$m width are analyzed every 0.005~$\mu$m, obtaining the regions of higher and lower variability of the whole wavelength range,  indicated in red and blue respectively in Figure \ref{fig:residuals}. With these regions, we define three spectral indices, applying the Equation \ref{eq:index}: 

\begin{equation}
    \textup{index} = \frac{\int_{\lambda_1}^{\lambda_2}F_1(\lambda)d\lambda}{\int_{\lambda_3}^{\lambda_4}F_2(\lambda)d\lambda}
    \label{eq:index}
\end{equation}

Where the highest flux-region, $F_{1}(\lambda)$ is integrated over the wavelength range in which the residuals are higher. $F_{2}(\lambda)$ is the integrated flux over the contiguous area over the wavelength range in which the residuals are closer to 0 (contiguous area of minimum variability).  We design three new spectral indices, as defined in Table \ref{Limits_ranges_index}, and we add three spectral indices that have been implemented in \cite{burgasser2006method} and \cite{Bardalez-Gagliuffi2014} and which we found helpful to find the flux maximum spectra. We have a total of six spectral indices.

Finally, we apply these indices to all 66 spectra of LP~261-75b, to produce a set of index-index plots as shown in Figure \ref{fig:index-index}. At a glance, we observe that the flux maximum spectra (associated with the outermost part of the light curve in Figure \ref{fig:LC_lp261}) cluster at different areas in the index-index plots than the flux maximum spectra (associated with the values closest to one light curve in the Figure \ref{fig:LC_lp261}). In the upcoming Section, we will objectively define the \textit{variable} and \textit{non-variable} areas within the index-index plots using a Machine Learning method.

\begin{deluxetable*}{ccccc}
\tablecaption{Limiting ranges used for the creation of spectral indices. References: (1) This document, (2) \citet{burgasser2006hubble} (3) \citet{gagliuffi2014spex} \label{Limits_ranges_index}}
\tablehead{
\colhead{Spectral} & \colhead{Numerator Range} & \colhead{Denominator Range} & \colhead{Feature}  &   \colhead{Reference}        \\           \colhead{index}    &     \colhead{($\mu$m)}    & \colhead{($\mu$m)}          &                  & 
}
\startdata
$J$-index             & 1.22-1.25        & 1.31-1.335      & Variability in $J$-band     &     (1)                    \\
$H$-index             &  1.64-1.67       & 1.545-1.575     & Variability in $H$-band     &        (1)                            \\
max-var               & 1.66-1.69        & 1.34-1.37       & Major difference variability in complete spectra  & (1)   \\
H$_2$O-J              & 1.14–1.165       &   1.26–1.285    & 1.15 $\mu$m H$_2$O          & (2)   \\
CH$_4$-J              & 1.315 - 1.335    & 1.26–1.285      & 1.32 $\mu$m CH$_4$          &      (2)                                \\
$J$-curve             & 1.26-1.29        & 1.14-1.17       & Curvature across  $J$-band  & (3)  
\enddata
\end{deluxetable*}

\subsection{Defining variability areas}

\begin{figure*}[h]
    \centering
    \includegraphics[width=0.85\linewidth]{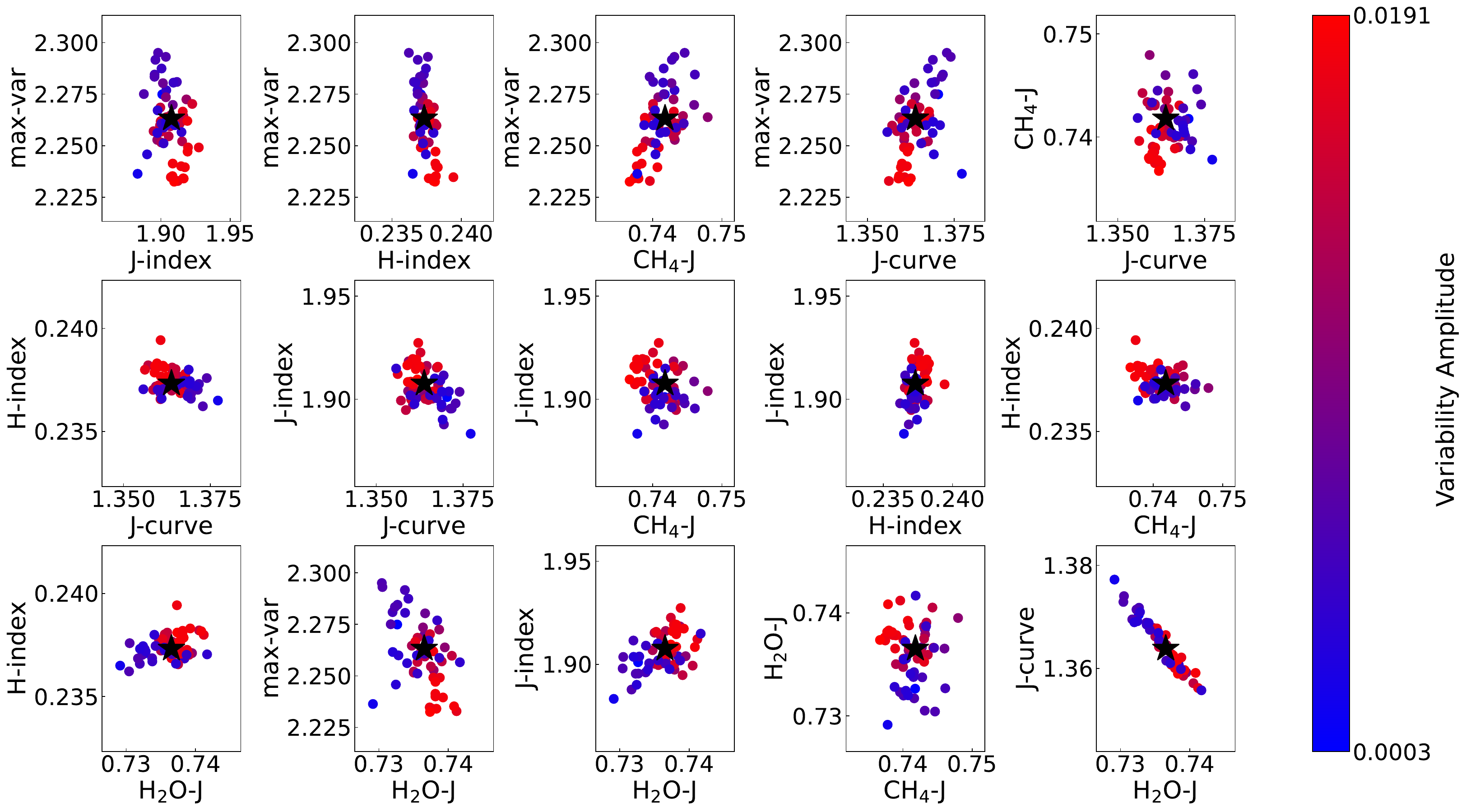}
    \caption{Index-index plots comparing all spectral indices for the HST/WFC3 spectra of LP~261-75b. The blue dots correspond to the spectra showing the smallest variability amplitude ($\sim$ 1.0 on the light curve in Figure \ref{fig:LC_lp261}), and the red dots correspond to the higher variability amplitude (about 1.01 or 0.98 on the light curve in Figure \ref{fig:LC_lp261}). The black star in the center corresponds to the calculated indices for the template spectrum, as a reference of a single-epoch spectrum.}
    \label{fig:index-index}
\end{figure*}

Given that we have a limited amount of spectra for LP~261-75b, before we could robustly define the \textit{variable} and \textit{non-variable} areas in the index-index plots in Figure \ref{fig:index-index}, we generated synthetic spectra, similar to the most and flux minimum spectra of LP~261-75b, to further help in the definition of both areas. This is the same method that we followed in \cite{oliveros2022informed}. 

\begin{figure*}[t]
    \centering
    \includegraphics[width=0.94\linewidth]{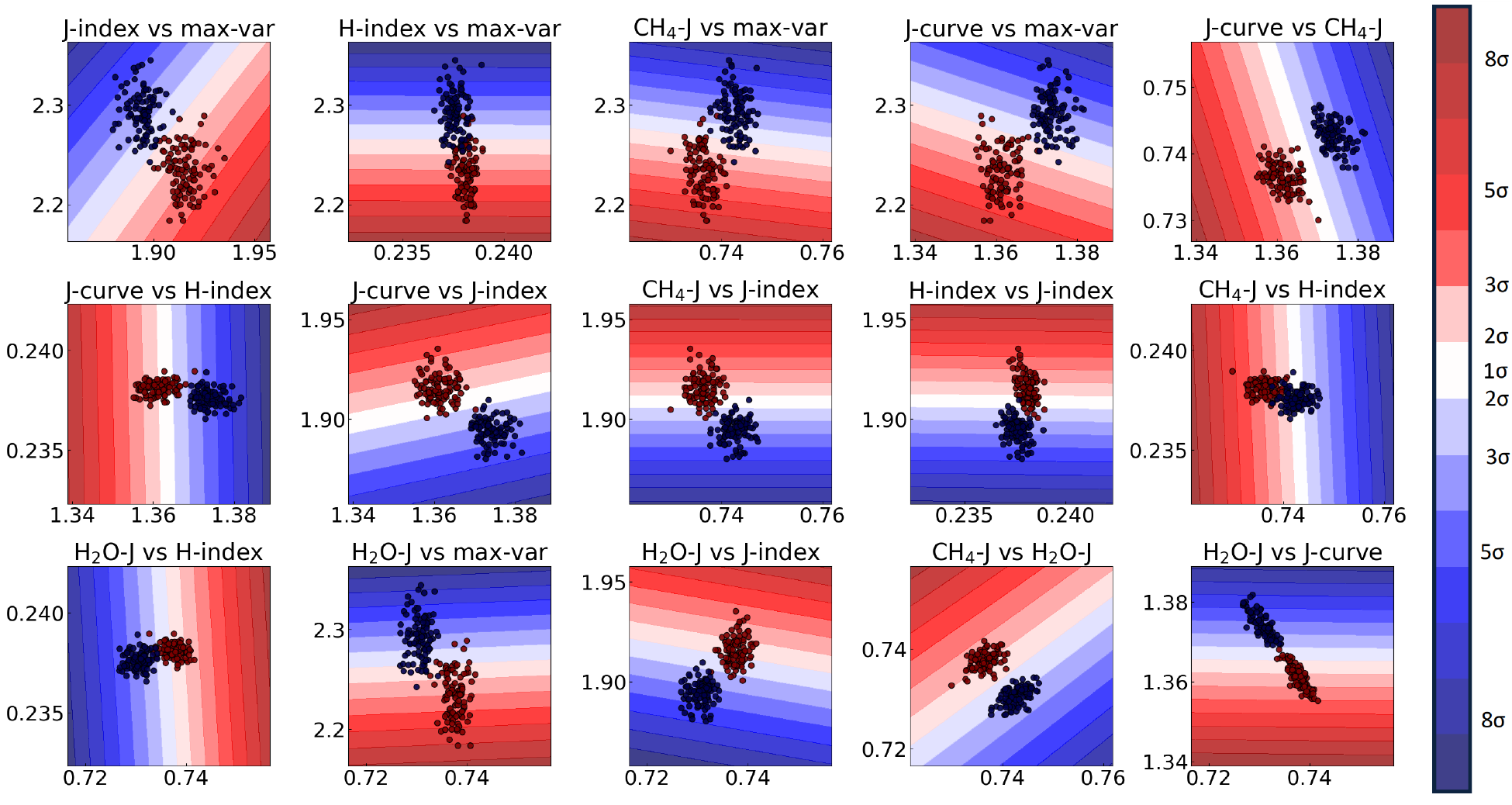}
    \caption{Index-index plot same as Figure \ref{fig:index-index} using the synthetic spectra generated with the Monte Carlo method from the HST/WFC3 spectra of LP~261-75b. The flux maximum spectra are shown as red dots, and the flux minimum spectra are shown as blue dots. We show the areas of variability computed using \textit{Machine learning} (\texttt{scikit-learn}) with their respective reliability regions.}
    \label{fig:index_regions}
\end{figure*}

We used a Monte Carlo simulation we generated 100 synthetic spectra from the flux minimum spectra of LP~261-75b, and 100 synthetic spectra from the flux maximum spectra. To generate the synthetic spectra, we selected the maximum and minimum flux spectra in the light curve and redefined each of their points in the spectrum using a Gaussian random number generator. The mean value of the Gaussian is the original flux of the spectrum, and the standard deviation is the uncertainty of each point in the sample spectrum. Using this method, we generated 100 synthetic spectra similar to the flux maximum spectra of LP~261-75b, and 100 similar to the flux minimum spectra. Finally, we ran our spectral indices through all synthetic spectra, and we generated new index-index plots (see Figure \ref{fig:index_regions}).

To robustly define the variable and non-variable areas of the index-index plots, we use a supervised machine learning method, \textit{Support Vector Machine} (SVM). To use the SVM method, we must provide a sample containing the points and their value that classifies it to one category or another. We input a sample with the value of each index-index plot for the 200 synthetic spectra created with the Monte Carlo simulation described above, and with the value assigned to it if it is variable (1) or non-variable (0).

The SVM method uses a portion of the sample values as training examples so that it can classify the two groups and evaluate them. This is in order to maximize the gap between the two categories. We use the rest of the sample data to map the space that predicts which category they belong to and evaluate the method.

We used \texttt{scikit-learn} to apply this supervised learning method \citep{scikit-learn} to our data. In addition, we need to select a kernel to apply in our method, to define the forms of pattern the analysis algorithms. Thus, we did kernel variations (linear, radial basis, and polynomial) to define the separation boundaries of the two categories. We decided to use a `linear' kernel because it proposed better-delimited boundaries between the two data sets. We did several tests varying the number and positions of the contour regions to confirm the reliability of finding an object in the variable or non-variable category, until our results stabilize, which determines the sigma value at which our method is reliable.

We note in Figure \ref{fig:index_regions} that the contours separating the two categories are not reliable to classify an object as variable or non-variable ($1-2\sigma$). Thus, we declare objects in the $3\sigma$ regions and above as being in a reliable variable or non-variable area. \textit{Variable areas} are shown in red $>3\sigma$, and \textit{non-variable areas} are shown in blue ($>3\sigma$).

Once we define the variable and non-variable areas, we notice with a visual inspection it is possible to observe that some indices seem to have more relevance than others. Therefore, in the Appendix (Section \ref{appendix}), we add some tests to track which combination of indices works best to find the variable and non-variable synthetic spectra.

\begin{figure}[h]
    \centering
    \includegraphics[width=0.8\linewidth]{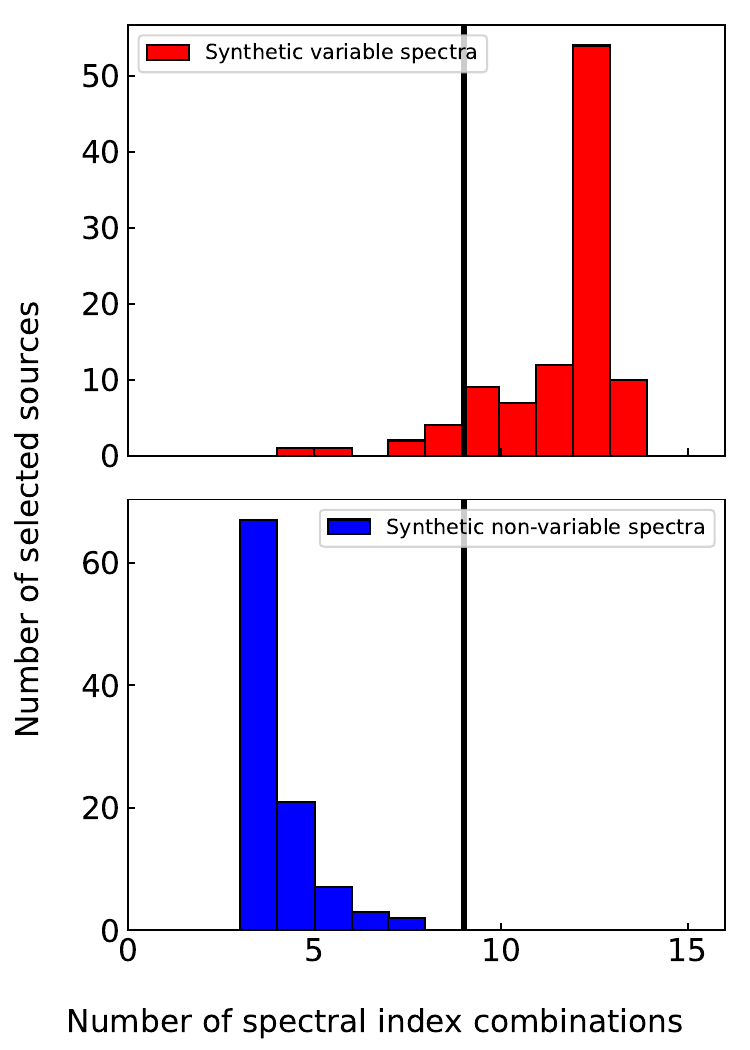}
    \caption{Histogram showing how many of the synthetically generated variable and non-variable spectra are found in the \textit{variable} or \textit{non-variable} areas. We use these histograms to define the minimum number of \textit{variable} areas inside the index-index plots in which one spectrum needs to be found to be considered \textit{variable} (threshold). Similarly for the \textit{non-variable} spectra.}
    \label{fig:histogram}
\end{figure}

In addition, we generate a histogram that includes the values of the indices calculated for all 200 spectra. It gives us information on how many of the 15 variable areas, each of our spectra is found in (see Figure \ref{fig:histogram}).

We define a variability threshold for the number of index-index plots in which a brown dwarf must be found in the variable zone to be considered a variable candidate. This threshold is 9 of 15 index-index plots. We select this number because for less than 9 plots 100\% of the non-variable synthetic spectra are in \textit{non-variable} areas (see in Figure \ref{fig:histogram}). In addition, in this limit, we obtain 87\% of the synthetic variable spectra found in at least 9/15 variable areas. Thus, any brown dwarf that is located in more than 9 index-index plots in the variable areas is considered a variable candidate. The spectra found in less than 9 index-index plots inside \textit{variable} areas are considered a non-variable candidate. 

The Python code to find L4--L8 brown dwarf variable candidates using the spectral indices presented in this work is publically available and ready to be used in Github: \url{https://github.com/ntlucia/BrownDwarf-SpectralIndices}.

\section{IDENTIFICATION OF VARIABLE CANDIDATES}\label{Sec:Candidate-variables}

Once robust \textit{variable} and \textit{non-variable} areas were defined within the index-index plots, we used our spectral indices method to identify variable L4--L8 brown dwarf candidates. 
Since the spectral indices presented here were designed with an L6 dwarf spectrum, these indices apply only to brown dwarfs with two spectral types difference (L4-L8). We applied our indices to a sample of 75 brown dwarfs with a single epoch near-infrared (0.7--2.5~$\mu$m) spectrum, and spectral type L4-L8 available at the SpeX/IRTF spectral library\footnote{\url{https://cass.ucsd.edu/~ajb/browndwarfs/spexprism/html/ldwarf.html}} (R$\sim$75-120).

\begin{figure}[h]
    \centering
    \includegraphics[width=0.9\linewidth]{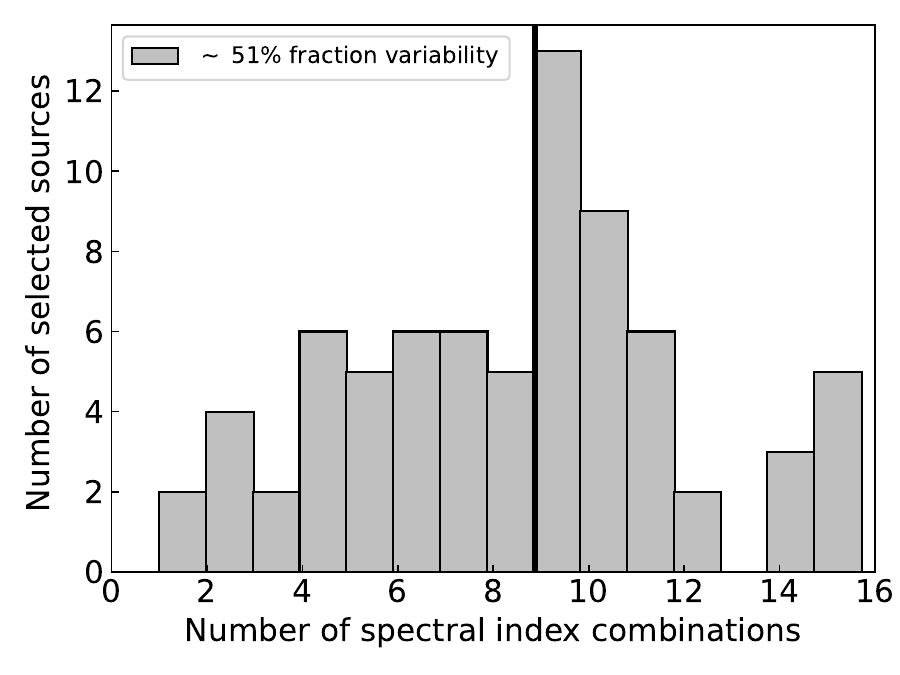}
    \caption{Histogram applying the spectral index method to the sample of 75 brown dwarfs, indicating the number of objects found in the variable regions.}
    \label{fig:hist_BD}
\end{figure}

After running our spectral indices on all of them, we concluded that 38 of the 75  L4--L8 brown dwarfs were found as variable candidates (found in 9 or more variable areas). This represents a $\sim 51\%$ variability fraction in our L4-L8 sample. In Figure \ref{fig:hist_BD}, we show the histogram with the number of index-index plots in which each of the objects is found in the \textit{variable} area.

From our sample, 23 of the 75 objects have been monitored photometrically in the literature, of which 11 are found to be variable (Section \ref{sec:variables}), 9 are found as non-variables in the literature and 3 are found as false negatives (Section \ref{sec:non-variables}). In the following section, we describe the objects selected as variable and non-variable candidates by our indices. The full list of variable candidates selected by our indices can be found in Table \ref{table:sample_testing}.

Including the known variable and non-variable objects that we mention in Table \ref{table:sample_testing}, we reproduce a similar analysis of the relevance of the combination of indices. In the Appendix (Section \ref{appendix}), we perform some tests to determine which index-index combinations are more effective in finding variable and non-variable L dwarfs.

\startlongtable
\begin{deluxetable*}{lccccccc}
\centering
\tablecaption{Sample of L dwarfs with spectral type between L4-L8 available at the SpeX Spectral Library. }
\label{table:sample_testing}
\tablehead{
\colhead{Name}      & \colhead{SpeX NIR}    & \colhead{Variable} & \colhead{N°} &  \colhead{Variability} &  \colhead{Variability}& & \colhead{Reference} \\
  & \colhead{SpT} &   \colhead{candidate?} & \colhead{regions} &   \colhead{literature$^a$} & \colhead{$J$-band [\%]}          & \colhead{3.6 $\mu m$ [\%]} &  \colhead{variability}
}
\startdata
L  362-29    B            & L5.5  & Yes & 15/15 &                   &      &      &                   \\
2MASS  J00282091+2249050  & L7    & No  & 05/15 &                   &      &      &                   \\
2MASSW  J0030300-145033   & L7    & No  & 04/15 & Variable*         &      & 1.52 & (1),  (2)         \\
2MASS  J00325937+1410371  & L8    & No  & 05/15 &                   &      &      &                   \\
LSPM  J0036+1821          & L4    & Yes & 09/15 & Variable          & 1.22 &    & (3)               \\
2MASSI  J0103320+193536   & L6    & No  & 03/15 & Slow  variability &      &   0.56    & (4),  (3)         \\
2MASSI  J0131183+380155   & L4    & Yes & 09/15 &                   &      &      &                   \\
2MASS  J01443536-0716142  & L5.5  & Yes & 10/15 &                   &      &      &                   \\
2MASS  J01550354+0950003  & L5    & Yes & 09/15 &                   &      &      &                   \\
DENIS  J020529.0-115925   & L7    & Yes & 11/15 &                   &      &      &                   \\
2MASSW  J0208236+273740   & L5    & Yes & 15/15 &                   &      &      &                   \\
2MASS  J02425693+2123204  & L4    & No  & 06/15 &                   &      &      &                   \\
2MASS  J02572581-3105523  & L8    & Yes & 09/15 &                   &      &      &                   \\
2MASSI  J0318540-342129   & L7    & Yes & 09/15 &                   &      &      &                   \\
2MASS  J04390101-2353083  & L6    & Yes & 14/15 & Variable          & 2.6  &      & (2)               \\
2MASS  J04474307-1936045  & L5    & Yes & 09/15 &                   &      &      &                   \\
2MASS  J06244595-4521548  & L5    & Yes & 11/15 & Variable          & 1    &      & (5),  (3)         \\
2MASS  J06523073+4710348  & L4.5  & No  & 04/15 &                   &      &      &                   \\
2MASS  J06540564+6528051  & L6    & Yes & 10/15 &                   &      &      &                   \\
2MASS  J07171626+5705430  & L6.5  & Yes & 10/15 &                   &      &      &                   \\
2MASSW  J0801405+462850   & L6.5  & No  & 8/15  & Non-Variable      &      &      & (5)               \\
2MASS  J08053189+4812330  & L4+T5 & Yes & 09/15 & Variable          & 1.46 & 3.13 & (6)               \\
2MASSW  J0820299+450031   & L5    & No  & 04/15 & Slow  variability &   &  $<0.4$ & (4),  (3)         \\
2MASSI  J0825196+211552   & L6    & Yes & 09/15 & Variable          & 1    & 1.4  & (5),  (4)         \\
2MASS  J08350622+1953050  & L4.5  & Yes & 11/15 &                   &      &      &                   \\
2MASS  J08354256-0819237  & L5    & Yes & 10/15 & Variable          & 1.7  &      & (2)               \\
2MASS  J08511627+1817302  & L5.5  & No  & 07/15 &                   &      &      &                   \\
2MASS  J08575849+5708514  & L8    & No  & 02/15 &                   &      &      &                   \\
2MASS  J09054654+5623117  & L5    & No  & 08/15 &                   &      &      &                   \\
2MASS  J09083803+5032088  & L7    & No  & 08/15 & Non-Variable      &      &      & (5)               \\
2MASS  J09153413+0422045  & L7    & Yes & 12/15 &                   &      &      &                   \\
2MASSW  J0929336+342952   & L7.5  & No  & 04/15 &                   &      &      &                   \\
2MASS  J10071185+1930563  & L8    & No  & 05/15 &                   &      &      &                   \\
2MASS  J10101480-0406499  & L6    & No  & 06/15 & Variable*         & 5.1  &      & (3),  (9)         \\
2MASSI  J1043075+222523   & L8    & No  & 02/15 &                   &      &      &                   \\
2MASS  J10433508+1213149  & L7    & Yes & 14/15 & Variable          &      & 1.54 & (3)               \\
2MASS  J10440942+0429376  & L7    & No  & 06/15 &                   &      &      &                   \\
2MASS  J11040127+1959217  & L4    & Yes & 09/15 &                   &      &      &                   \\
HD 97334B                 & L4.5  & No  & 6/15  &                   &      &      &                   \\
2MASS  J11211858+4332464  & L7.5  & Yes & 14/15 &                   &      &      &                   \\
DENIS  J112639.9-500355   & L4.5  & Yes & 09/15 & Variable          & 1.2  & 0.21 & (2),  (4)         \\
2MASS  J11555389+0559577  & L7    & No  & 06/15 & Non-variable      &      &      & (7)               \\
2MASS  J12195156+3128497  & L8    & No  & 06/15 & Non-variable      &      &      & (5),  (3)         \\
SIPS  J1228-1547          & L5.5  & Yes & 12/15 &                   &      &      &                   \\
2MASSW  J1239272+551537   & L5    & No  & 05/15 &                   &      &      &                   \\
2MASS  J13262981-0038314  & L5.5  & No  & 01/15 &                   &      &      &                   \\
2MASS  J14002320+4338222  & L7    & Yes & 15/15 &                   &      &      &                   \\
ULAS2MASS  J1407+1241     & L5    & Yes & 15/15 &                   &      &      &                   \\
2MASS  J14222720+2215575  & L6.5  & Yes & 09/15 &                   &      &      &                   \\
2MASS  J14283132+5923354  & L4    & Yes & 10/15 &                   &      &      &                   \\
2MASSW  J1507476-162738   & L5.5  & Yes & 10/15 & Variable          &      & 0.53 & (4),  (3)         \\
2MASSW  J1515008+484742   & L6.5  & No  & 04/15 & Non-variable      &      &      & (5)               \\
2MASS  J15150607+4436483  & L7.5  & No  & 07/15 &                   &      &      &                   \\
2MASS  J15200224-4422419B & L4.5  & Yes & 12/15 &                   &      &      &                   \\
Gl  584c                  & L8    & No  & 01/15 &                   &      &      &                   \\
2MASSI  J1526140+204341   & L7    & Yes & 10/15 &                   &      &      &                   \\
2MASS  J16303054+4344032  & L7.5  & No  & 02/15 &                   &      &      &                   \\
2MASS  J16335933-0640552  & L6    & No  & 07/15 &                   &      &      &                   \\
2MASS  J17054834-0516462  & L4    & Yes & 10/15 &                   &      &      &                   \\
2MASSI  J1711457+223204   & L6.5  & No  & 08/15 &                   &      &      &                   \\
2MASS  J17461199+5034036  & L5    & Yes & 11/15 &                   &      &      &                   \\
2MASS  J17502484-0016151  & L5.5  & Yes & 10/15 & Variable          & 1.12 &      & (2),  (5),    (3) \\
2MASS  J18212815+1414010  & L5    & Yes & 11/15 & Variable          &      & 0.54 & (4)               \\
2MASS  J19285196-4356256  & L5    & Yes & 11/15 &                   &      &      &                   \\
2MASSI  J2002507-052152   & L6    & No  & 07/15 &                   &      &      &                   \\
2MASSW  J2101154+175658   & L7.5  & No  & 02/15 &                   &      &      &                   \\
2MASS  J21321145+1341584  & L6    & Yes & 09/15 &                   &      &      &                   \\
2MASSW  J2148162+400359   & L6    & No  & 03/15 & Slow  variability &      & 1.33 & (4),  (3)         \\
2MASSI  J2158045-155009   & L4    & Yes & 09/15 &                   &      &      &                   \\
2MASS  J22120703+3430351  & L5    & No  & 04/15 &                   &      &      &                   \\
2MASSW  J2224438-015852   & L4.5  & No  & 07/15 & Non-variable      &      &      & (4),  (3)         \\
2MASS  J22443167+2043433  & L6.5  & No  & 06/15 & Variable*         & 5.5  & 0.8  & (7),  (3)         \\
2MASS  J22521073-1730134  & L7.5  & No  & 07/15 &                   &      &      &                   \\
2MASSI  J2325453+425148   & L8    & No  & 05/15 &                   &      &      & 
\enddata
\begin{tablenotes}
 \small
\item \textbf{Notes:} $^a$ \textit{Variable} refers to the objects that present any measure variability in the literature,  \textit{Non-Variable} refers to the objects that no measure variability in photospectroscopic studies. \textit{slow variability} refers to the objects that discuss the small variability, in some cases not consistent or near to the RMS in the measurements. Variable* refers to the false negative in our sample because with our method we obtain non-variable objects.   \textbf{Variability references: }(1) \cite{clarke2008search}, (2) \cite{radigan2014strong}, (3) \cite{vos2020spitzer}, (4) \citet{metchev2015weather}, (5) \cite{buenzli2014brown}, (6) \cite{dupuy2012hawaii}, (7) \cite{koen2004search}, (8) \cite{vos2018variability}, (9) \cite{wilson2014brown}.
\end{tablenotes}
\end{deluxetable*}

\subsection{Variables candidates}\label{sec:variables}

The following objects were selected as variable candidates by our indices and have information about their variability in the literature.

\subsubsection{LSPM J0036+1821}
J0036+1821 is an L4 dwarf \citep{geballe2002toward}. It is located at $8.7356 \pm 0.0105$ pc \citep{2020GaiaEDR3}. \citet{gagne2015banyan} found an effective temperature of $\mathrm{T_{eff}} =1900$~K, and a surface gravity of log(g) = 5.0~dex by fitting the Spectral Energy Distribution (SED) of the object using the BT-Settl atmospheric models. \cite{osorio2005optical} was the first to report variability of this object in an $I$-band of 0.25\%. Later, \cite{metchev2015weather} used J0036+1821 as a control object for recognizing potential activity-induced photometric effects. They found a variability amplitude in the [3.6] Spitzer channel of $0.47 \pm 0.05 \%$, and a variability amplitude of $0.19 \pm 0.05 \%$ in the [4.5] Spitzer channel. They measured a period of $2.7 \pm 0.3$ hr. \cite{metchev2015weather} did not include it as a variable in their final sample since its variability might be associated with magnetic activity. \cite{croll2016longterm} measured a variability amplitude of $A_J = 1.22 \pm 0.04 \%$. J0036+1821 was found in 9 out of 15 areas in our index–index plots, indicating that it is a variable candidate.

\subsubsection{2MASS J04390101-2353083}
J0439-2553 is an L4.5 dwarf \citep{schneider2014discovery}. It is located at a distance of  $12.387 \pm 0.0550$ pc \citep{2020GaiaEDR3}. \cite{gagne2015banyan} found an effective temperature of $\mathrm{T_{eff}} =1600$~K, and a surface gravity of log(g) = 5.0~dex by fitting the SED of the object using the BT-Settl atmospheric models.

\cite{wilson2014brown} found a variability amplitude for J0439-2553 of $2.6 \pm 0.5 \%$ and  \cite{radigan2014independent} found $<$1.2 \% using the same SofI instrument. J0439-2553 was found in the variable areas in 9 out of 15 of our index–index plots, indicating it is a variable candidate.

\subsubsection{2MASS J06244595-4521548}
J0624-4521 is an L6.5 dwarf \citep{schneider2014discovery}.
It is located at a distance of $12.1914 \pm 0.0532$ pc \citep{2020GaiaEDR3}. \cite{gagne2015banyan} found an effective temperature of $\mathrm{T_{eff}} =1500$~K, and a surface gravity of log(g) = 5.0~dex by fitting the SED of the object using the BT-Settl atmospheric models.

\cite{buenzli2014brown} measured variability of at least 1\%, using observations very short, 40 minutes, so they identified variability but did not measure amplitude and period very precisely. J0624-4521 was found in the variable areas in 11 out of 15 of our index–index plots, indicating that it is a variable candidate.

\subsubsection{2MASS J08053189+4812330}
J0805+4812 was a spectral binary candidate \citep{2007bburgasser}. It was confirmed as such by \cite{dupuy2012hawaii}. They show large perturbations due to orbital motion with spectral types L4 (primary) + T5  (secondary). J0805+4812 shows variability in their near-IR light curve associated with the transits binary system and not necessarily internal processes of the atmospheres such as clouds \citep{dupuy2012hawaii}. Although this system is binary, we use for our analysis the resolved spectrum of the object, the primary dwarf L4.  

J0805+4812 was found in the variable areas in 9 out of 15 of our index–index plots, indicating that it is a variable candidate.

\subsubsection{2MASSI J0825196+211552}
J0825+2115 is an L7.5V dwarf \citep{kirkpatrick200067}. \cite{gagne2015banyan} found an effective temperature of $\mathrm{T_{eff}} =1500$~K, and a surface gravity of log(g) = 5.0~dex by fitting the SED of the object using the BT-Settl atmospheric models.

\cite{buenzli2014brown} measured significant variability in the $J$- and $H$-band ($A_{J,H}>1 \%$). \cite{metchev2015weather} found a variability amplitude of $A_{[3.6]} =0.81 \pm 0.08 \%, A_{[4.5]} 1.4 \pm 0.3 \%$, in Spitzer, with a period of 7.6 hr. J0825+2115 was found in 9 out of 15 variable areas of our index–index plots, indicating that it is a variable candidate.

 \subsubsection{2MASS J08354256-0819237}
J0835+1953 is an L5 dwarf \citep{schneider2014discovery}. It is located at a distance of  $26.1 \pm 5.1$ pc \citep{schmidt2010colors}. \cite{gagne2015banyan} found an effective temperature of $\mathrm{T_{eff}} =1600$~K, and a surface gravity of log(g) = 5.0~dex by fitting the SED of the object using the BT-Settl atmospheric models.

\cite{radigan2014independent} measured a peak-to-peak variability of 1.3 $\pm$ 0.2\% using the SofI instrument, installed on the New Technology Telescope (NTT) in $J$-band. J0835+1953 was found in the variable areas in 11 out of 15 of our index–index plots, indicating that it is a variable candidate.

\subsubsection{2MASS J10433508+1213149}
J1043+1213 is an L7 dwarf \citep{burgasser2010spex}. It is located at a distance of  $18.2 \pm 3.8 $ pc \citep{schmidt2010colors}. \citet{gagne2015banyan}  found an effective temperature of $\mathrm{T_{eff}} =1500$~K, and a surface gravity of log(g) = 5.5~dex by fitting the SED of the object using the BT-Settl atmospheric models. \cite{metchev2015weather} measured a variability amplitude of $A_{[3.6]} = 1.54 \pm 0.15 \%$, $A_{[4.5]} = 1.2 \pm 0.2$ \%, with a period $3.8 \pm 0.2$ hr, but irregular, i. e., the period fits require more Fourier terms than the number of recorded rotations. J1043+1213 was found in the variable areas in 14 out of 15 of our index–index plots, indicating that it is a variable candidate.

\subsubsection{DENIS J112639.9-500355}
J1126-5003 is an L5.5 dwarf \citep{phan2008discovery}. It is located at a distance of $16.1726 \pm 0.0654$~pc \citep{2020GaiaEDR3}. \cite{folkes2007discovery} reported that J1126-5003 has unusually blue colors for its L4.5 optical or L6.5 NIR spectral type. \cite{radigan2014strong} measured a peak-to-peak variability amplitude in the $J$-band of $1.2 \pm 0.1 \%$. \cite{metchev2015weather} measured  a variability amplitude of $A_{[3.6]} = 0.21 \pm 0.04 \%$, $A_{[4.5]} = 0.29 \pm 0.15\%$, and they calculated a period of $3.2 \pm 0.3$ hr with regular periodicity. J1126-5003 was found in the variable areas in 9 out of 15 of our index–index plots, indicating that it is a variable candidate.

\subsubsection{2MASSW J1507476-162738}
J1507-1627 is an L5V dwarf \citep{kirkpatrick200067}. It is located at a distance of $7.4102 \pm 0.0143$ pc \citep{2020GaiaEDR3}. \citet{gagne2015banyan} found an effective temperature of $\mathrm{T_{eff}} =1900$~K, and a surface gravity of log(g) = 5.0~dex by fitting the SED of the object using the BT-Settl atmospheric models. 

\cite{metchev2015weather} measured a variability amplitude of $A_{[3.6]} = 0.53 \pm 0.11\%$, $A_{[4.5]} = 0.45 \pm 0.09\%$, with an irregular period of $2.5 \pm 0.1$ hr. 
\cite{metchev2015weather} explains that J1507-1627 shows clear spot evolution, since one oscillation appears around 7hr in one channel observing sequence, and continuously grows in amplitude until the end of the other channel sequence, 2.5~hr rotations later. J1507-1627 was found in the variable areas in 10 out of 15 of our index–index plots, indicating that it is a variable candidate.

\subsubsection{2MASS J17502484-0016151}
J1750-0016 is L5.5 dwarf \citep{burgasser2010spex}. It is located at a distance of $9.2097 \pm 0.0168 $ pc \citep{2020GaiaEDR3}. \cite{gagne2015banyan} found an effective temperature of $\mathrm{T_{eff}} =1800$~K, and a surface gravity of log(g) = 4.5~dex by fitting the SED of the object using the BT-Settl atmospheric models. 

\cite{buenzli2014brown} measured a significant variability in the $J$-band, between 1.12 and 1.32 $\mu$m, with a period $>$ 4hr. However, the variation in the $H$-band is not statistically significant. J1750-0016 was found in the variable areas in 10 out of 15 of our index–index plots, indicating that it is a variable candidate.

\subsubsection{2MASS J18212815+1414010}
J1821+1414 is an L5 dwarf \citep{schneider2014discovery}. \cite{gagne2015banyan} found an effective temperature of $\mathrm{T_{eff}} =1800$~K, and a surface gravity of log(g) = 4.5~dex by fitting the SED of the object using the BT-Settl atmospheric models. \cite{looper2008discovery} classified this object as peculiarly red indicating moderately low gravity, including relative weakness in the alkaline and FeH strengths and sharpness of the $H$-band continuum. \cite{gagne2014banyan} concluded that despite its signatures of youth this object does not belong to any known young moving group \citep{faherty2016population}. 

\cite{metchev2015weather} measured a variability amplitude of $A_{[3.6]} = 0.54 \pm 0.05$ \%, $A_{[4.5]} = 0.71 \pm 0.14$ \%, and an irregular period of $4.2 \pm 0.1$ hr. This object is an example of variability in objects with peculiar red spectra or low gravity objects, which are interesting in our sample and are in agreement with other results of spectral indices such as \cite{ashraf2022disentangling} and studied with more detail in \cite{Vos2022}.
J1821+1414 was found in the variable areas in 11 out of 15 of our index–index plots, indicating that it is a variable candidate.

\subsection{Non-variables candidates}\label{sec:non-variables}
In this Section, we list the objects that were selected by our indices as non-variable candidates and are listed as such in the literature.

\subsubsection{2MASSI J0103320+193536}
J0103+1935 is an L6V \citep{kirkpatrick200067} dwarf. It is located at a distance of $23 \pm 2$ pc \citep{faherty2009brown}. \cite{gagne2015banyan} found an effective temperature of $\mathrm{T_{eff}} =1600$~K, and a surface gravity of log(g) = 5.5~dex by fitting the SED of the object using the BT-Settl atmospheric models.

\cite{metchev2015weather} measured a small variability amplitude of $A_{[3.6]} = 0.56 \pm 0.03$\% and $A_{[4.5]}	= 0.87 \pm 0.09$\%. J0103+1935 was found in the variable areas in 3 out of 15 of our index–index plots, indicating that it is a non-variable candidate. Even though J0103+1935 shows some variability, our index method is probably not sensitive to such small variability levels.

\subsubsection{2MASSW J0801405+462850}
J0801+4628 is an L6.5V dwarf \citep{kirkpatrick200067}. It is located at a distance of $19.1 \pm 3.7$ pc \citep{schmidt2010colors}. \cite{gagne2015banyan} found an effective temperature of $\mathrm{T_{eff}} =1500$~K, and a surface gravity of log(g) = 5.0~dex by fitting the SED of the object using the BT-Settl atmospheric models. 

\cite{buenzli2014brown} did not find any significant variability above their uncertainties. J0801+4628 was found in the variable areas in 8 out of 15 of our index–index plots, indicating that it is a non-variable candidate. 

\subsubsection{2MASSW J0820299+450031}
J0820+4500 is an L5V dwarf \citep{kirkpatrick200067}, located at a distance of $42.7 \pm 8.4$ pc. \cite{metchev2015weather} measured variability $A_{[3.6]} < 0.4$\% and $A_{[4.5]}	< 0.48$\%,  classifying it as non-variable. J0820+4500 was found in the variable areas in 4 out of 15 of our index–index plots, indicating that it is a non-variable candidate. 

\subsubsection{2MASS J09083803+5032088}
J0908+5032 is an L8 dwarf \citep{schneider2014discovery}. It is located at a distance of $10.4244 \pm  0.0596$ pc. \cite{gagne2015banyan} found an effective temperature of $\mathrm{T_{eff}} =1600$~K, and a surface gravity of log(g) = 5.5~dex by fitting the SED of the object using the BT-Settl atmospheric models. 

\cite{buenzli2014brown} did not measure any significant variability above their uncertainties using HST/WFC3 time-resolved spectroscopy. J0908+5032 was found only in 8 out of 15 variable areas of our index–index plots, indicating that it is a non-variable candidate. 

\subsubsection{2MASS J11555389+0559577}

J1155+0559 is an L7.5 dwarf \citep{knapp2004near}. It is located at a distance of $17.27 \pm 3.04$ pc \citep{faherty2012brown}. This object is somewhat controversial and interesting concerning its spectral type and variability, for which they have even cataloged it as  L6-L8 peculiar \citep{Gagne2015}. \cite{gizis2002palomar} noted the presence of $H\alpha$ emission which can be associated with magnetic fields variability. In addition,  \cite{koen2003search} 
monitored the object for 5 nights of observation, of about 4.6 hr each, in $I$-band, observing variability and obtaining a period of 8~hr.

\cite{koen2004search} monitored J1155+0559 in the $K$- and $H$-bands finding variability with a periodicity of 45~min, which is an extremely short rotational period for a brown dwarf. In this study, J1155+0559 was found in the variable areas in 6 out of 15 of our index–index plots, indicating that it is a non-variable candidate. 

\subsubsection{2MASS J12195156+3128497}
J1219+3128 is an L8 dwarf \citep{schneider2014discovery}. It is located at a distance of $18.1 \pm 3.7$ pc \citep{schmidt2010colors}. \citet{tannock2021weather} found an effective temperature of $\mathrm{T_{eff}} =1330$~K, and a surface gravity of log(g) = 5.1~dex by fitting the SED of the object using the BT-Settl and SM08 atmospheric models. 

\cite{buenzli2014brown} measured a ramp in the 1.12-1.2 $\mu$m wavelength range, with a small degree of variability, and they presented J1219+3128 as a tentative variable. J1219+3128 was found in the variable areas in 6 out of 15 of our index–index plots, indicating that it is a non-variable candidate. 

\subsubsection{2MASSW J1515008+484742}
J1515+4847 is an L6.5 dwarf \citep{cruz2003meeting}. It is located at a distance of $9.8073 \pm 0.0246$ pc \citep{2020GaiaEDR3}.  \cite{gagne2015banyan} found an effective temperature of $\mathrm{T_{eff}} =1600$~K, and a surface gravity of log(g) = 5.0~dex by fitting the SED of the object using the BT-Settl atmospheric models. 

\cite{buenzli2014brown} found no evidence of significant variability in any of the spectral regions for J1515+4847, showing less than 0.5-1\% variability per hour, which is within the observational uncertainties. J1515+4847 was found in the variable areas in 4 out of 15 of our index–index plots, indicating that it is a non-variable candidate. 

\subsubsection{2MASSW J2148162+400359}

J2148+4003 is an L7 dwarf \citep{schneider2014discovery} at a distance of $8.0857 \pm 0.0235$ pc \citep{2020GaiaEDR3}, and it is a very red brown dwarf \citep{looper2008discovery}. \cite{metchev2015weather} measured variability of $A_{[3.6]} = 1.33 \pm 0.07$\%, $A_{[4.5]} = 1.03 \pm 0.10$\%, but in a long observation window (14h). They classify J2148+4003 as non-variable because they only detect flux fluctuations over a long time. J2148+4003 was found in the variable areas in 3 out of 15 of our index–index plots, indicating that it is a non-variable candidate.

\subsubsection{2MASSW J2224438-015852}
J2224-0158 is an L4.5V dwarf \citep{kirkpatrick200067}. It is located at a distance of $11.6075 \pm 0.0549$ pc \citep{2020GaiaEDR3}. \cite{burningham2021cloud} found an effective temperature of $\mathrm{T_{eff}} =1912$~K, and a surface gravity of log(g) = 5.47~dex using atmospheric retrievals.

\cite{metchev2015weather} did not find variability on this object. \cite{gelino2002dwarf} did not measure any variability in the $I$-band for J2224-0158. In addition, \cite{burningham2021cloud} showed us that it has enstatite and quartz clouds, but that the body may have an unfavorable geometry to detect rotational modulation signals, or that the dark regions are arranged latitudinally in bands, and this may be why variability in near-infrared wavelengths is not observed. J2224-0158 was found in the variable areas in 6 out of 15 of our index–index plots, indicating that it is a non-variable candidate.

\subsubsection{False Negatives\label{sec:false-negative}}

In this section, we describe the objects that we determine as false negatives because our spectral index method determines them as non-variable objects, but in the literature, they have been measured to have variability in different spectral bands.

\textit{2MASSW J0030300-145033:}
J0030-1450 is an L7 dwarf \citep{kirkpatrick200067}. It is located at a distance of $26.7 \pm 3.3$ pc \citep{faherty2009brown}. \cite{gagne2015banyan} found an effective temperature of $\mathrm{T_{eff}} =1500$~K, and a surface gravity of log(g) = 5.0~dex by fitting the SED of the object using the BT-Settl atmospheric models. 

\cite{enoch2003photometric}  detected periodic variability in $K_s$-band, with an amplitude of 0.19~mag and period of 1.5~hr. However \cite{clarke2008search} found no evidence of variability, with amplitude smaller than 40~mmag in $J$-band. Similarly, \cite{radigan2014strong} observed a flat light curve for this target in the $J$-band throughout 3.2~hr. \cite{vos2022let} found variability of $A_{[3.5 \mu m]} = 1.52 \pm 0.06 \%$ using Spitzer and obtained a period of	$4.22 \pm 0.02$ hr.  Therefore, despite discrepancies among the literature results, they are not mutually exclusive. As the observations are not simultaneous, this may be an example where the extent of variability varies over time. J0030-1450 was found in the variable areas in 5 out of 15 of our index–index plots, indicating that it is a non-variable candidate. This may be a false negative of our sample.

\textit{2MASS J10101480-0406499:}
J1010-0406 is field L6 dwarf \citep{cruz2003meeting, reid2008meeting}. It is located at a distance of $18 \pm 2$ pc. \cite{gagne2015banyan} found an effective temperature of $\mathrm{T_{eff}} =1600$~K, and a surface gravity of log(g) = 5.0~dex by fitting the SED of the object using the BT-Settl atmospheric models. 

\cite{wilson2014brown} found a variability amplitude of  $5.1 \pm 1 \%$ in an observation window of only 3~hr, using the SofI instrument ($J-$band). They did not report a period, because the window observation is too short for a proper determination. J1010-0406 was found in the variable areas in 6 out of 15 of our index–index plots, indicating that it is a non-variable candidate. This is a false negative object in our sample.

\textit{2MASS J22443167+2043433:}
J2244+2043 is an L6.5gamma \citep{faherty2016population}, young and  low-gravity dwarf \citep{gizis2015wisep}. It is located at a distance of $19 \pm 2$ pc \citep{faherty2009brown}. \cite{faherty2016population} found an effective temperature of $\mathrm{T_{eff}} =1184$~K.

\cite{morales2006sensitive} and \cite{vos2018variability} found variability very close to the RMS amplitude, with Spitzer at 4.5 $\mu$m of observation and no variability at 3.6 $\mu$m during 5.7 hr and $> 8$ hr, respectively. In addition,  \cite{vos2018variability} found no variability in $J$-band during 4hr using UKIRT/WFCAM but they predict with the periodogram a period of $P > 5.5$~hr.

This suggests that the light curves of this object may have evolved in time or that they possess longitudinal bands with sinusoidal surface brightness modulations and an elliptical spot. When two bands have slightly different periods due to different velocities or directions, they interfere to produce a beating between patterns, according to \cite{apai2017zones}. J2244+2043 was found in the variable areas in 6 out of 15 of our index–index plots, indicating that it is a non-variable candidate.

\subsection{New Variable Candidates}\label{sec:new_variable_candidates}

The following objects listed in Table \ref{table:new-variables} are selected as new variable candidates by our spectral indices, not reported before in the literature.

\startlongtable
\begin{deluxetable*}{ccccccccc}
\centering
\tablecaption{New variable candidates calculated with our indices, no reports of variability in the literature.}
\label{table:new-variables}
\tablehead{
\colhead{Name}               &  \colhead{SpeX NIR}    & \colhead{Variable} & \colhead{N°} &  \colhead{Distance} & \colhead{$T_{eff}$} & \colhead{$\log g$}& \colhead{References} & \colhead{Additional}\\
  & \colhead{SpT} &   \colhead{candidate?} & \colhead{regions} &   \colhead{[pc]} & \colhead{[K]} & \colhead{[dex]} &   & \colhead{notes}
}
\startdata
L 362-29 B                    & L5.5                              & Yes                                 & 15/15                            & 12.1438 $\pm$ 0.0376   & 1600  & 5         & (1), (2)                    & (a)                       \\
2MASSI J0131183+380155        & L4                                & Yes                                 & 09/15                            & 24.4358 $\pm$ 0.2956   &                &           &     (1)                               &                                                      \\
2MASS J01443536-0716142       & L5.5                              & Yes                                 & 10/15                            & 12.7392   $\pm$ 0.0804 & 1600    & 5         & (1), (2)               &                                                      \\
2MASS J01550354+0950003       & L5                                & Yes                                 & 09/15                            & 22.2014 $\pm$ 0.2687                & 1800           & 4.5       &                        (1)            &                                                      \\
DENIS J020529.0-115925        & L7V                                & Yes                                 & 11/15                            & 20.2649 $\pm$ 0.6684   & 1500           & 5.5       &    (1), (2)                 &                                       \\
2MASSW J0208236+273740        & L5V                                & Yes                                 & 15/15                            & 24 $\pm$ 2             & 1600  & 5         & (3), (2)                    &                                                      \\
2MASS J02572581-3105523       & L8                                & Yes                                 & 09/15                            & 9.7371 $\pm$ 0.0451    &                &           &                                    &                                                      \\
2MASSI J0318540-342129        & L7                                & Yes                                 & 09/15                            & 14.0882 $\pm$ 0.3668   & 1600  & 5         & (1), (2)                    &                                                      \\
2MASS J04474307-1936045       & L5                                & Yes                                 & 09/15                            & 26 $\pm$ 5             &                &           & (3), (2)                   &                                                      \\
2MASS J06540564+6528051       & L6                                & Yes                                 & 10/15                            & 38.2 $\pm$ 7.5         & 1800           & 5         & (4), (2) &                                                      \\
2MASS J07171626+5705430       & L6.5                              & Yes                                 & 10/15                            & 28.946 $\pm$ 0.4877    & 1700           & 5         & (1), (2)  &                                                      \\
2MASS J08354256-0819237       & L5                                & Yes                                 & 10/15                            & 7.2298  $\pm$ 0.0112   & 1800           & 5         & (1),(2)  &                                                      \\
2MASS J09153413+0422045       & L7                                & Yes                                 & 12/15                            & 18.0 $\pm$ 4.2         & 1600  & 5.5       & (5), (2)          &                                                      \\
2MASS J11040127+1959217       & L4                                & Yes                                 & 09/15                            & 17.952 $\pm$ 0.1245    & 1900           & 5         & (1), (2)                    &                                                      \\
2MASS J11211858+4332464       & L7.5                              & Yes                                 & 14/15                            & 46 $\pm$ 6             &                &           & (3)                            &                                                      \\
SIPS J1228-1547               & L5.5                              & Yes                                 & 12/15                            & 21.84$\pm$ 1.8         & 1700           & 5         & (1), (2)                    &                                                      \\
2MASS J14002320+4338222       & L7                                & Yes                                 & 15/15                            & 23.7 $\pm$ 11.1        & 1600           & 5.5       & (3),(4)                  &                                                      \\
ULAS2MASS J1407+1241          & L5                                & Yes                                 & 15/15                            & 28.4074 $\pm$ 1.2877   & 1600           & 5.5       &    (1), (2)                                &                                                      \\
2MASS J14222720+2215575       & L6.5                              & Yes                                 & 09/15                            & 31.4 $\pm$ 6.4         &                &           & (4)                  &                                                      \\
2MASS J14283132+5923354       & L4.4V                                & Yes                                 & 10/15                            & 21.949 $\pm$ 0.1819    & 1800           & 5.5       &   (1),(2)                                 &                                                      \\
2MASS J15200224-4422419B      & L4.5                              & Yes                                 & 12/15                            & 19 $\pm$ 1             &                &           & (3)                            & (b) \\
2MASSI J1526140+204341        & L7                                & Yes                                 & 10/15                            & 20.6983 $\pm$ 0.333    & 1600           & 5         &    (1),(2)                                &                                                      \\
2MASS J17054834-0516462       & L4                                & Yes                                 & 10/15                            & 18.825 $\pm$ 0.0852    & 2300           & 5.5       &    (1),(2)                                &                                                      \\
2MASS J17461199+5034036       & L5.7V                                & Yes                                 & 11/15                            & 20.685 $\pm$ 0.1914    & 1900           & 5         &     (1),(2)                                &                                                      \\
2MASS J19285196-4356256       & L5                                & Yes                                 & 11/15                            & 28.59 $\pm$ 0.8220     &                &           &     (1)                               &                                                      \\
2MASS J21321145+1341584       & L6                                & Yes                                 & 09/15                            & 27.77 $\pm$ 0.7        & 1500           & 5.5       &          (1),(2)                           &                                                      \\
2MASSI J2158045-155009        & L4                                & Yes                                 & 09/15                            & 23.197 $\pm$ 0.4883    &                &           &       (1)                             &              
\enddata
\begin{tablenotes}
 \small
\item \textbf{References} (The first one in the column is the distance and the second one is the physical parameters): (1) \cite{2020GaiaEDR3}, (2) \cite{gagne2015banyan}: BT-Settl models, (3) \cite{faherty2009brown}, (4) \cite{schmidt2010colors}, (5) \cite{reid2008meeting}. \textbf{Additional notes}: (a) well-separated binary system, (b) well-separated (1.174'' $\pm$ 0.016'') binary system.\\
\end{tablenotes}
\end{deluxetable*}

\begin{figure*}
    \centering
    \includegraphics[width=1\linewidth]{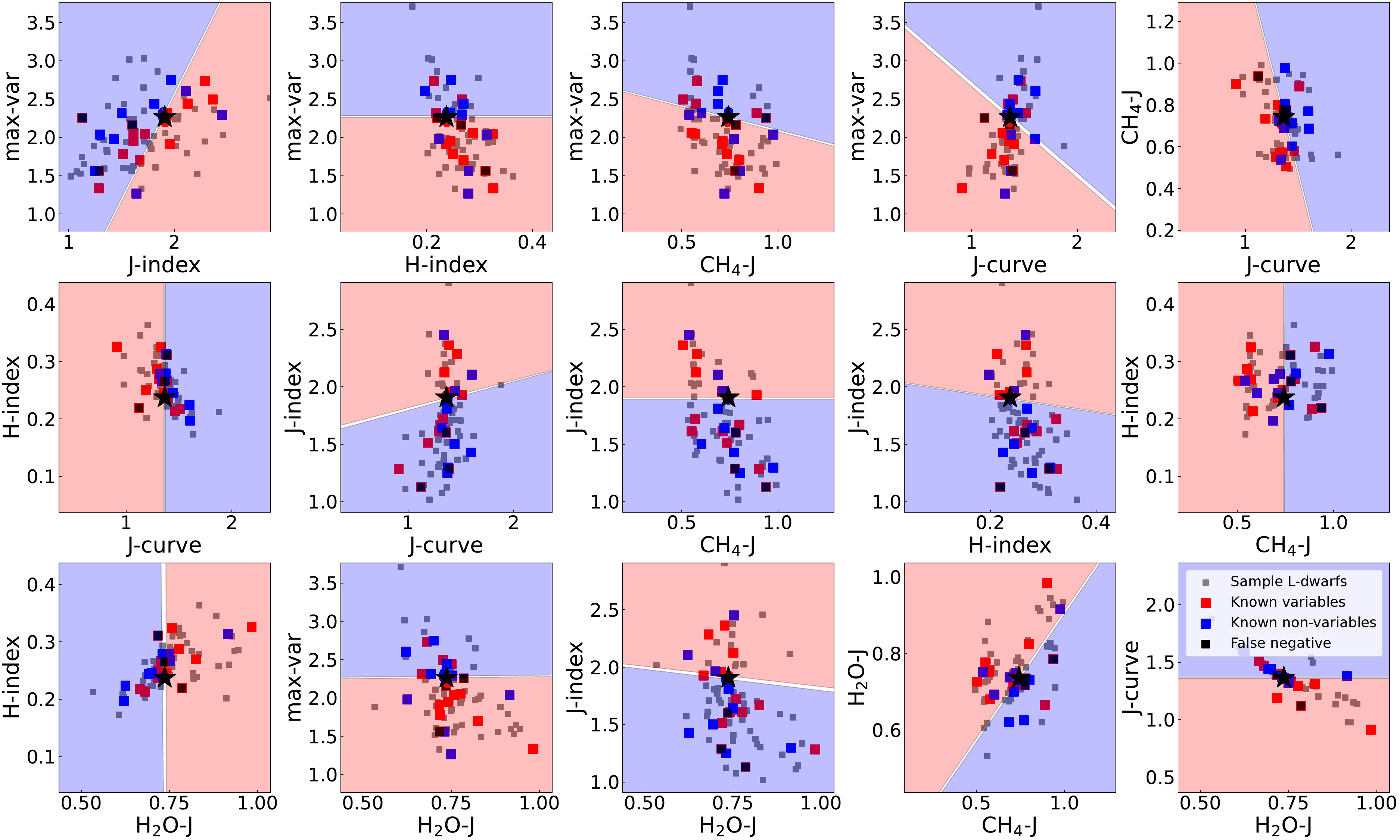}
    \caption{Index-index plots to our sample in (grey squares). In addition, the known variables (red squares), non-variable objects in our sample (blue squares), and the false negatives (black squares). The red and blue colored areas of variability and non-variability, respectively, are explained in Section \ref{sec:index-method} (sections \ref{sec:variables} and \ref{sec:non-variables}).}
    \label{fig:index-index-sample}
\end{figure*}

In Figure \ref{fig:index-index-sample} we show the distribution in the index-index plots of all the objects in our sample, differentiating between the known variables, non-variables, and false positives (Section \ref{sec:variables} and \ref{sec:non-variables}). Also, in Table \ref{table:sample_testing}, we show the information of the L4-L8 sample with IRTF/SpeX spectra.

\section{Discussion}\label{Sec:discussion}

In this Section, we discuss the results obtained and the relevance for future brown dwarf variability studies.

\subsection{Recovery Rates}\label{recovery_rates}

From the sample of 75 L4--L8 brown dwarfs with IRTF/SpeX spectra, a sub-sample of 23 brown dwarfs was monitored for variability studies in the literature (see Table \ref{table:sample_testing}). From those 23 objects, our results show 11 are variable candidates (see Section \ref{sec:variables}), and the other 12 are non-variables candidates (see \ref{sec:non-variables}). From the sample of 75 L4--L8 brown dwarfs with IRTF/SpeX spectra, a sub-sample of 23 brown dwarfs was found with variability studies in the literature (see Table \ref{table:sample_testing}). From those 23 objects, our results show 11 are variable candidates (see Section \ref{sec:variables}), and the other 12 are non-variables candidates (see \ref{sec:non-variables}). For the 11 variable candidates, the literature reports all of them as variables. Thus, our spectral indices were able to recover 100\% of all the variable brown dwarfs, demonstrating the tool's ability to find the most likely variable brown dwarfs.

Of the 12 non-variable candidates, 9 of them are consistent as non-variable in the literature. But, we obtained three of them, J0030-1450, J0103+1935, and J1010-0406 \citep{cruz2003meeting}, were classified as non-variable by our spectral indices as it was found in a \textit{variable} area in less of 9 out of 15 index-index plots and in the literature measure variability in some band of near-infrared, or mid-infrared. Therefore, the recovery rate for variable L4--L8 brown dwarfs for the spectral indices presented here is $\sim 80\%$, also we obtain the false negative rate as $\sim$25\% (3/12 objects).

This is particularly important to conduct successful rotational modulation campaigns of brown dwarfs, in particular in this spectral range (L4--L8) in which not all brown dwarfs show significant photometric or spectro-photometric variability (see Section \ref{comparison_surveys}). In this work, we provide the list of the most-likely variable L4--L8 brown dwarfs according to our spectral indices in Section \ref{sec:new_variable_candidates}, and the list of L4--L8 brown dwarfs to avoid in monitoring campaigns in Section \ref{sec:variables} and\ref{sec:non-variables}.

\subsection{Limit of variability detection}\label{sec:limit-variability }

\begin{figure}[h]
    \centering
    \includegraphics[width=0.9\linewidth]{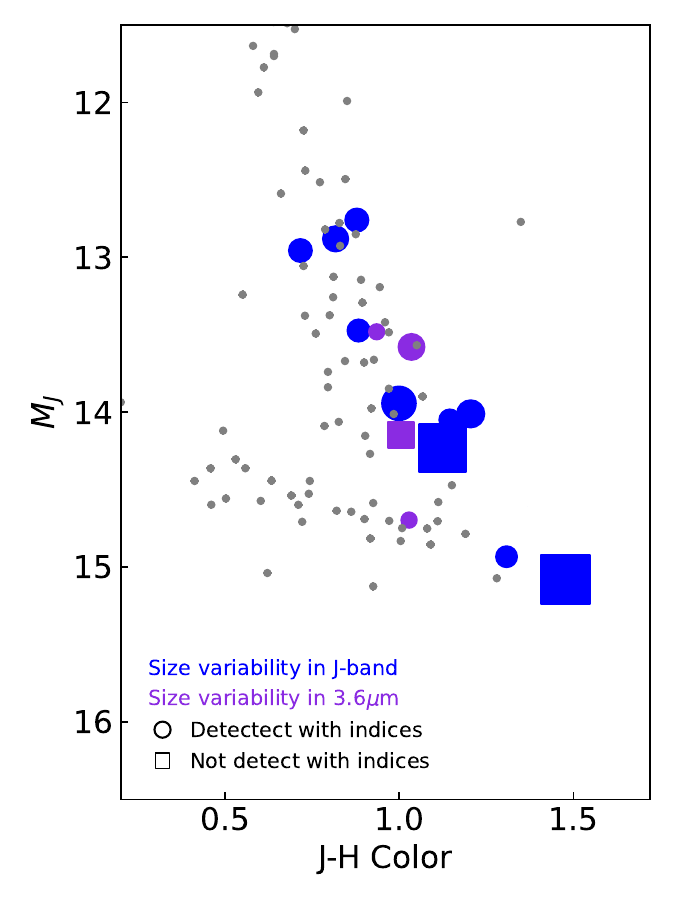}
    \caption{Color-magnitude diagram of brown dwarfs in the MKO  system. The grey dots are the showing all ultracool dwarfs from Table 3 in \citet{dupuy2012hawaii}. The colors represent the variability in $J$-band (blue) or 3.6 $\mu$m (purple), and the size represents $\%$ of variability. Also, the squares are the false negative objects.}
    \label{fig:CM-variables}
\end{figure}

In Figure \ref{fig:CM-variables}, we show a color-magnitude plot of all variable objects in the literature, where we differentiate the percentages of $J$-band and 3.6 $\mu$m variability, to identify the variability limit that our indices can detect. In Figure \ref{fig:CM-variables}, we observed how two of the three objects that we report as false negatives (J1010-0406 and J2244+2043), in the literature have variability with values much higher than normally reported. The variability of these objects was measured with ground-based instruments, which are less reliable light curves, since they are more likely contaminated by clouds, telluric absorptions from the atmosphere, etc. This contamination does not happen in light curves obtained with space-based observatories. For the false positives, the variability status has changed with time (See discussion in Section \ref{sec:false-negative}). Thus, for this study, we discard these measurements and define the variability lower limit in 1\% for space telescope measurements in 3.6 $\mu$m, 4.5 $\mu$m, and $J$-band. However, since for ground-based observations, the variability that is detected depends on the observatory, the telescope size, and the meteorological conditions, we do not define a lower limit for this type of variability reporting. 

Additionally, it is worth noting that the 3 false negatives shown in Figure \ref{fig:CM-variables} are redder in the $J$-band and of a later type than the bulk of the variables. Although the number of objects we have, is a small number statistic, this may point to a lower effectiveness of our method for later or low-severity objects.  However, perhaps this is something we can study in detail in the future.

\subsection{Variability Fraction}\label{comparison_surveys}

We estimate the variability fraction in the L4--L8 spectral range using the spectral indices introduced in Section \ref{sec:index-method} over the sample of 75 L4--L8 brown dwarfs in the IRTF/SpeX spectral library, assuming that in fact, all brown dwarfs flagged as variables by our indices are indeed variable. We compare the variability fraction obtained with our index method to the fraction obtained by other brown dwarf variability surveys in the literature.

In this work, we found that 12 out of the 75 are known variables, flagged as variables by our indices (Section \ref{sec:variables}). In addition, our indices found 27 new variable candidates for which no variability information was available in the literature (Section \ref{sec:new_variable_candidates}). In total, our indices found 38 L4--L8 variable candidates out of the 75 L4--L8 brown dwarfs with IRTF/SpeX spectra (51\%). In the literature, only 11 of the 75 are confirmed by variability surveys as variable, bringing the minimum variability fraction to $\sim$14\% for L4--L8 dwarfs. As in \cite{oliveros2022informed}, we follow the same method as \cite{Burgasser2003} to estimate the maximum variability fraction using a Poisson distribution. We use as the mean of the distribution the percentage of variable candidates found by our indices (51\%), and the minimum variable fraction (14\%) with the variable objects confirmed in the literature as the minimum of the distribution. We integrated the probability until reaching 0.68 (1-$\sigma$ limit) which gives us the upper limit for the variability fraction. Following this procedure, the variability fraction estimated for L4--L8 dwarfs is $\mathrm{51^{+4}_{-38}}$\%.

As we did in \cite{oliveros2022informed} for mid- and late-T dwarfs, we compare the variability fraction that we obtained with our indices to that calculated by ground-based \citep{radigan2014strong} and space-based brown dwarf variability surveys \citep{buenzli2014brown, metchev2015weather}. \cite{radigan2014strong} performed a brown dwarf variability survey on 57 L, L/T, and T brown dwarfs using the Du Pont 2.5~m telescope at Las Campanas Observatory and the Canada–France–Hawaii Telescope on Maunakea. \cite{radigan2014strong} obtained a variability fraction of $60^{+22}_{-18}$\% for brown dwarfs outside the L/T transition (L9.0-T3.5), although they do not report a specific variability fraction for the L4--L8 spectral range, this fraction is consistent to that estimated in this work.

Finally, we compare our results to the variability fraction obtained by \cite{buenzli2014brown} and \cite{metchev2015weather} using the HST/WFC3 instrument and the Spitzer telescope, respectively. \cite{buenzli2014brown} monitored a sample of 22 L, L/T, and  T dwarfs, and found that four brown dwarfs in the range L5-L8 showed significant or tentative variability out of the 12 dwarfs observed in that spectral range ($\sim$33\%). \cite{metchev2015weather} used the [3.6] and [4.5] Spitzer channels, and monitored a sample of L, L/T, and T dwarfs, from which 16 were in the L4--L8 spectral range. Eight out of the 15 L4--L8 brown dwarfs were classified as a variable ($\sim$50\%).  After correcting for sensitivity, \cite{metchev2015weather} found that $\mathrm{80^{+20}_{-27}}$\%
of L dwarfs show variability amplitudes higher than 0.2\%. Within uncertainties, our variability fraction is in agreement with the variability fraction estimated by \cite{radigan2014strong}, \cite{buenzli2014brown} and \cite{metchev2015weather}.

\subsection{Physics behind the index method}\label{physics}

\begin{figure}
    \centering 
    \includegraphics[width=0.5\textwidth]{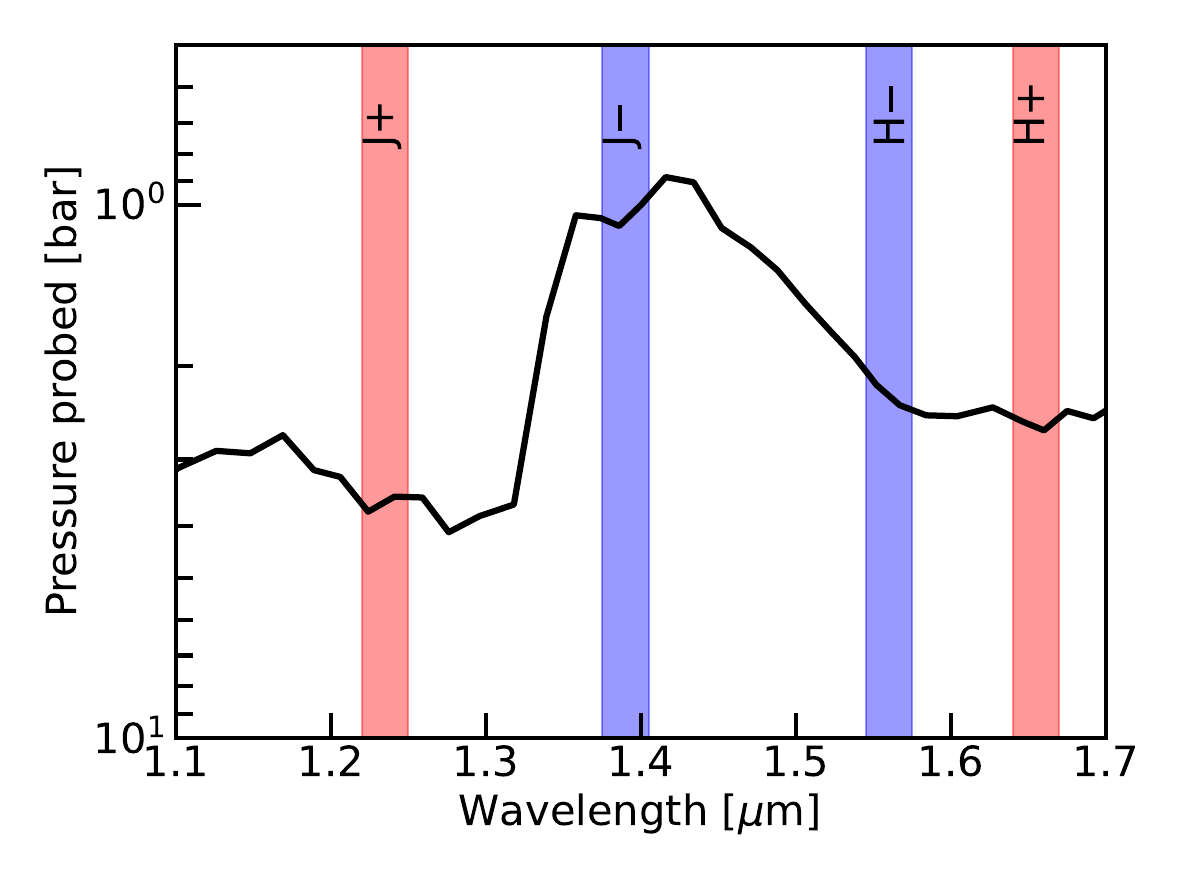}
    \caption{Pressure levels trace by the HST/WFC3 + G141 wavelength range in the atmosphere of LP~621-75b. The red (J+ and H+) and blue bands (J- and H-) are the most variable ranges, and least variable ranges, respectively, used in the design of the $J$- and $H$-index (see Table \ref{Limits_ranges_index}). }
    \label{fig:pressure_levels}
\end{figure}

\begin{figure}
    \centering
    \includegraphics[width=0.5\textwidth]{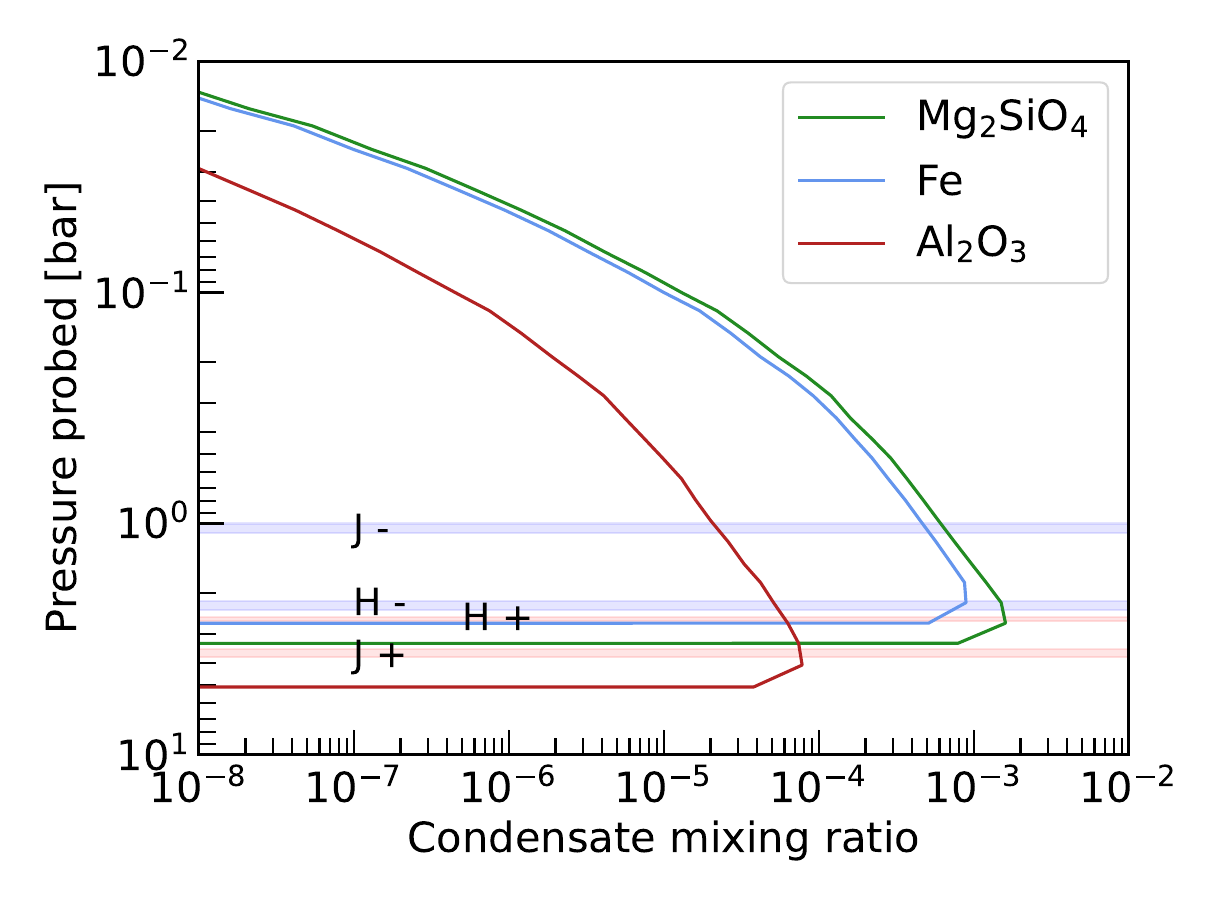}
    \caption{Condensate volume mixing ratio (mole fraction) in the model plotted as a function of model pressure. Where $+$ means the more variable region in the spectra used to design the index and $-$ is the less variable region in the spectra, similar to Fig. \ref{fig:pressure_levels}. We show the mole fraction of each cloud at each pressure level: $\mathrm{Mg_{2}SiO_{4}}$, green line, Fe, blue line, and $\mathrm{Al_{2}O_{3}}$, red line.}
    \label{fig:cloud_condensation}
\end{figure}

Our spectral indices compare the wavelength ranges of the HST/WFC3 spectrum of LP~261-75b which show the most variability, and the ranges that show the least as the object rotates (see Section \ref{sec:index-method}).  As shown in Figure \ref{fig:residuals} the wavelength ranges that vary the most are between 1.22 and 1.25~$\mu$m in $J$-band, and between 1.64 and 1.67~$\mu$m in the $H$-band, and those that vary the least are close to the water band at 1.4~$\mu$m (1.375-1.405~$\mu$m and 1.545-1.575~$\mu$m). To understand why certain wavelength ranges show higher flux variations than others, we use the contribution functions predicted for an object of similar effective temperature and surface gravity by radiative-transfer models \citep{Saumon_Marley2008}, and also the condensate mixing ratio (mole fraction) of each cloud expected for LP~261-75b ($\mathrm{Mg_{2}FeO_{4}}$, Fe and $\mathrm{Al_{2}O_{3}}$) shown in Fig. \ref{fig:cloud_condensation}. In Figure \ref{fig:pressure_levels}  we show the pressure levels probed by HST/WFC3 and the G141 grism for a field L6.0 brown like LP 261--75B ($\mathrm{T_{eff}}$ = 1500~K, log~g = 4.5, \citealt{bowler2013planets}) according to radiative-transfer models \citep{Saumon_Marley2008}. 

\begin{figure}
    \centering
    \includegraphics[width=0.45\textwidth]{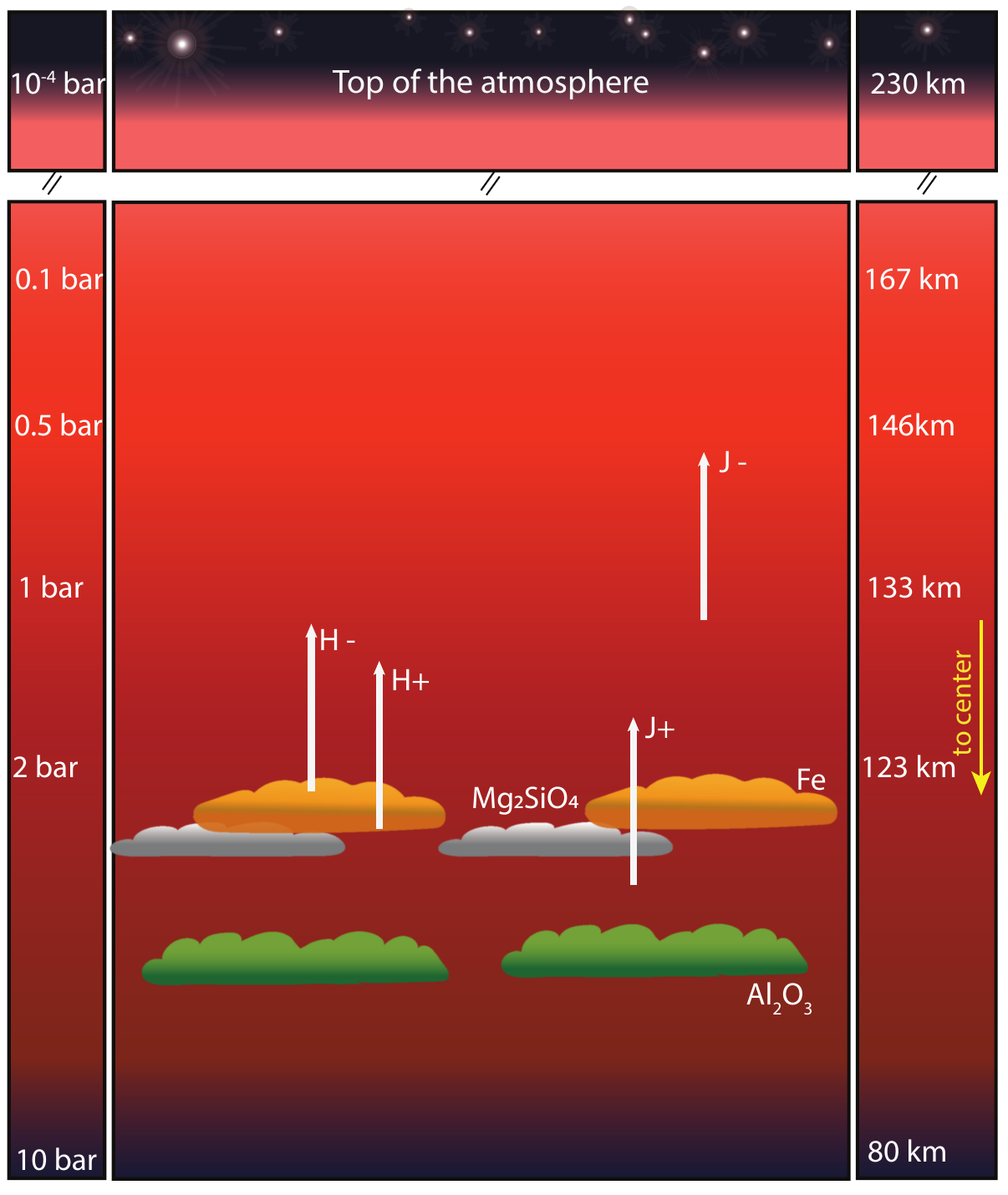}
    \caption{Representation of the vertical cloud structure of LP~261-75b and brown dwarfs of similar spectral types. We indicate where the Fe, $\mathrm{Mg_{2}SiO_{4}}$, and $\mathrm{Al_{2}O_{3}}$ clouds condensate, and the pressure levels trace by each wavelength range relevant for some of our spectral indices.}
    \label{fig:representation_clouds}
\end{figure}

As observed in Figure \ref{fig:pressure_levels}, the most variable wavelength ranges in the $J$- and $H$-bands (J+ and H+) trace higher pressure levels (deeper) in the atmospheres of mid-L dwarfs. The 1.22--1.25~$\mu$m region traces the atmosphere of LP~261-75b at 3.64~mbar, and the 1.64-1.67~$\mu$m region traces the atmosphere at 2.60~mbar. The least variable regions of the spectrum between 1.375-1.405~$\mu$m and 1.545-1.575~$\mu$m, trace the 1.04~mbar and 2.27~mbar pressure levels, respectively ($J$- and $H$-). The most variable areas in the $J$- and $H$-bands are tracing a thicker and deeper cloud layer which might introduce higher variability amplitudes in localized wavelength ranges of the HST/WFC3 spectra. On the other hand, the pressure levels traced by the least variable regions are more superficial in comparison, with less cloud coverage present at those atmospheric levels (see Figures \ref{fig:pressure_levels} and \ref{fig:cloud_condensation}), which might explain the lower variability measure on those ranges.  As observed in Figure \ref{fig:cloud_condensation}, and in Figure \ref{fig:representation_clouds} for a qualitative representation, the most variable regions of the $J$- and $H$-bands (J+ and H+) trace pressure levels with more cloud coverage, potentially explaining the higher variability measure on those wavelength ranges. Even though the H+ and H- regions trace relatively close pressure levels of LP 261-75B atmosphere, H- is tracing a pressure level with less mole fraction of the $\mathrm{Mg_{2}FeO_{4}}$ cloud likely explaining the smaller variability amplitude found in the H- wavelength range.

\subsection{Beyond Brown Dwarf Variability}

Brown dwarfs share colors and effective temperatures with some directly-imaged exoplanets \citep{faherty2016population}. Particularly in the mid-to late-L spectral type range, a handful of directly-imaged exoplanets are found with this spectral classification: $\beta$-Pictoris~b \citep{Lagrange2009}, with a $\mathrm{T_{eff}}$ = 1650±150 \citep{bonnefoy2014}. HIP~65426~b \citep{Chauvin2017}, with a $\mathrm{T_{eff}}$ = $\mathrm{1500^{+100}_{-200}}$~K \citep{Chauvin2017}. AB-Pictoris~b \citep{Chauvin2005}, with a $\mathrm{T_{eff}}$ = $2000^{+100}_{–300}$~K and log~g = 4.0±0.5 \citep{Bonnefoy2010}, and kap And b \citep{Carson2013} with a $\mathrm{T_{eff}}$ = $\mathrm{1900^{+100}_{-200}}$~K \citep{Bonnefoy2014b}.

Time-resolved spectroscopic data provides a wealth of information about the atmospheric dynamic and composition of brown dwarfs, which is most likely due to heterogeneous cloud structures at different pressure levels of the atmospheres of brown dwarfs that evolve within several rotations \citep{apai2017zones}. Using radiative-transfer models \citep{Saumon_Marley2008} we can infer which types of clouds are introducing the variability found at those pressure levels \citep{yang2016extrasolar, manjavacas2021revealing}. In addition, if we monitor the object spectroscopically during several rotations, we can infer the map of its surface, and reconstruct which atmospheric characteristics (spots and/or bands) are shaping the light curve produced at different wavelengths \citep{karalidi2015aeolus, Luger2019}. Time-resolved spectroscopic campaigns have been performed for several dozens of brown dwarfs (\citealt{buenzli2014brown, radigan2014strong, metchev2015weather, Biller2018, manjavacas2017cloud, vos2020spitzer}, among others), but only for one system of directly-imaged exoplanets HR~8799bc \citep{apai2016high,biller2021high} for which only upper levels of variability could be set, due to the low signal-to-noise of the data. 
Luckily, after the successful launch and commissioning of the \textit{James Webb Space Telescope}, we are obtaining unprecedented quality spectra for directly-imaged exoplanets. Spectroscopic variability of these objects provides extremely valuable information about the dynamics, clouds, and structure of directly-imaged exoplanets \citep{Hinkley2022, Patapis2022, Carter2023}. Nonetheless, to ensure the success of the first rotational monitoring campaigns for directly-imaged exoplanets, spectral indices like those presented in this work might be extremely helpful to pre-select the most likely candidates to show variability. Until now no systematic method to pre-identify variable L dwarfs was available, thus these spectral indices will very likely save hours of expensive telescope time.

\section{Conclusions}\label{Sec:conclusions}

\begin{enumerate}
\item We created novel spectral indices to identify L4--L8 variable candidate brown dwarfs using a single-epoch near-infrared spectrum. For designing these spectral indices, we employed time-resolved HST/WFC3 near-infrared spectra of LP~261-75b, one known variable L6 brown dwarf \citep{manjavacas2017cloud}.

\item We designed three new spectral indices to find potential L4--L8 variable brown dwarfs. In addition, we used three other spectral indices from previous studies \citep{burgasser2010spex, Bardalez-Gagliuffi2014}, to complete a total of six indices. We created 15 index-index plots which allow us to segregate variable and non-variable spectra of LP~261-75b. Using a Monte Carlo simulation of the flux maximum and flux minimum spectra of LP~261-75b we created 100 extra synthetic spectra of both types to clearly define the \textit{variable} and \textit{non-variable} areas in the index-index plots. Using a Support Vector Machines (SVM) supervised classification method, we objectively defined the \textit{variable} and \textit{non-variable} areas within each index-index plot.

\item We created a histogram showing how many \textit{variable} or \textit{non-variable} areas typically each of the variable or non-variable synthetic spectra are found. Using the histogram, we defined the variability threshold in 9/15 index-index plots. The spectra/objects found in more than 9 out of 15 variable areas are classified as \textit{variable candidates}. 

\item We applied our spectral indices to 75 L4--L8 IRTF/SpeX near-infrared spectra, which allowed us to identify potential mid-L variable brown dwarfs. Thirty-eight out of the 75 L4--L8 objects were classified as \textit{variable candidates} by our spectral indices. Eleven of these objects were reported as known variables in the literature. In addition, we found 27 new variable candidates. With these results, we found a variability fraction of ${51}_{-38}^{+4}$\% for our mid-L dwarf sample. Our variability fraction agrees with those provided by \cite{buenzli2014brown}, \cite{radigan2014strong}, and \cite{metchev2015weather} for the same range of spectral types.

\item From the 37 L4--L8 dwarfs classified as non-variable by our spectral indices, 12 were reported in the literature as non-variables. All of the non-variable objects are non-variable using our method, which represents the 100\% recovery rate in non-variable objects. This is crucial since this method will allow us to vet those L4--L8 brown dwarfs that most likely will not show variability in a monitoring campaign.

\item Three objects, J0030-1450, J0103+1935, and J1010-0406 were classified as non-variable by our spectral indices, but it was reported as a variable object in the literature. With these results, the false negative rate is $\sim$16\%.

\item We report 27 new L4--L8 variable candidates, which will need to be confirmed with near-infrared time-resolved photometry. 

\item Given that some L4--L8 dwarfs share effective temperatures and colors with some directly-imaged exoplanets, these spectral indices might be used in the future to find the most likely variable directly-imaged exoplanets to be monitored with the \textit{James Webb Space Telescopes} or the 30-m type upcoming ground-based telescopes.

\end{enumerate}

N. O. G. acknowledges the support from the Arthur Davidsen Graduate Student Research Fellowship \textit{provided by} the Space Telescope Science Institute.

J. M. V. acknowledges support from a Royal Society - Science Foundation Ireland University Research Fellowship (URF$\backslash$1$\backslash$221932).


\bibliography{sample631}{}
\bibliographystyle{aasjournal}



\section{Appendix }\label{appendix}
In our visual inspection of the Figures \ref{fig:index-index} and \ref{fig:index_regions}, we noticed that some index-index combinations seemed more important than others in finding variable or non-variable objects. Here we show which index-index combinations are more relevant than others.

\subsection{Relevance of each index-index combination}

To test which of the index-index combinations presented in Section \ref{sec:index-method} are more relevant to find variable or non-variable L dwarfs, we measured the number of variable and non-variable spectra that fall into each of the variability regions in all the possible combinations of the indices. We applied this methodology to both the variable and non-variable synthetic spectra shown in Figure \ref{fig:index_regions}, as well as to all known variable and non-variable objects in the literature (red and blue squares in Figure \ref{fig:index-index-sample}, respectively).

\begin{figure}[h]
    \centering
    \includegraphics[width=0.48\linewidth]{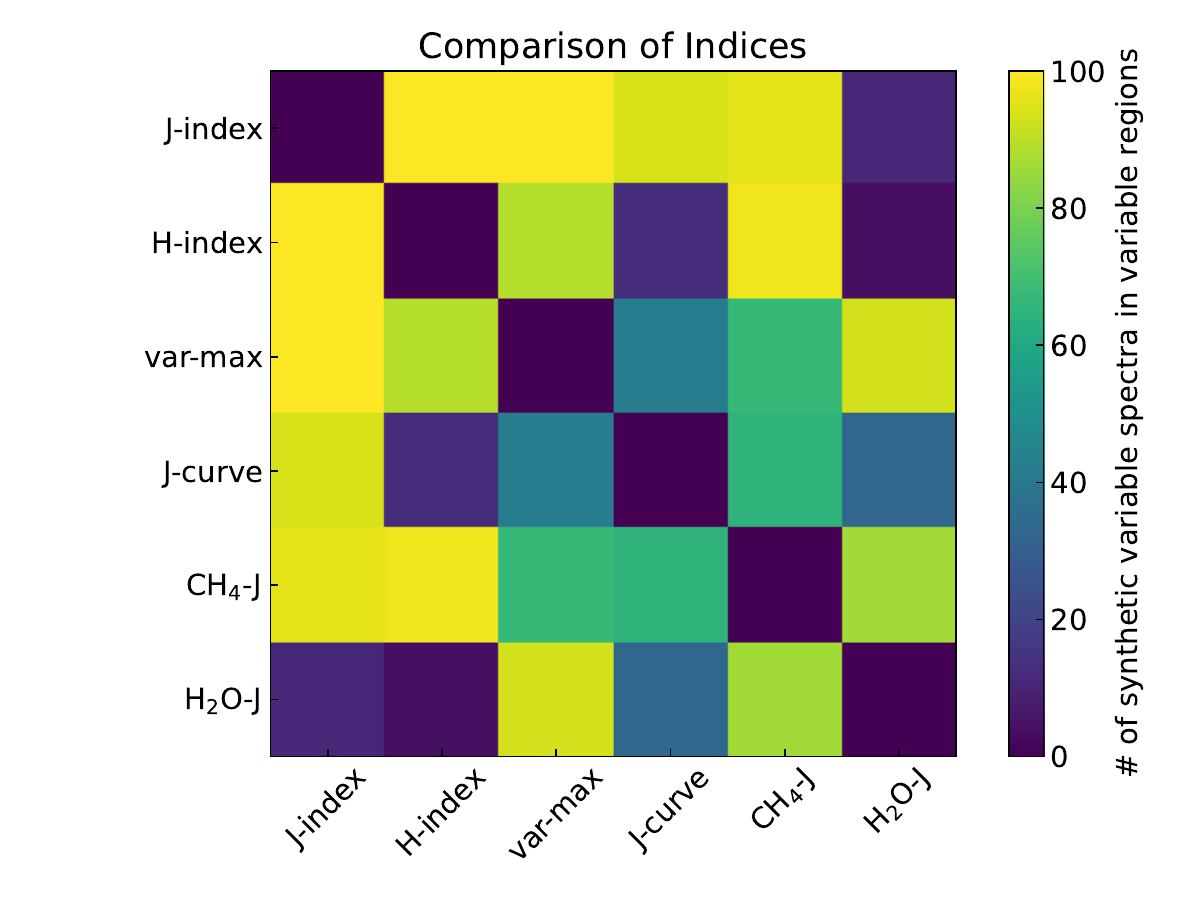}
    \includegraphics[width=0.48\linewidth]{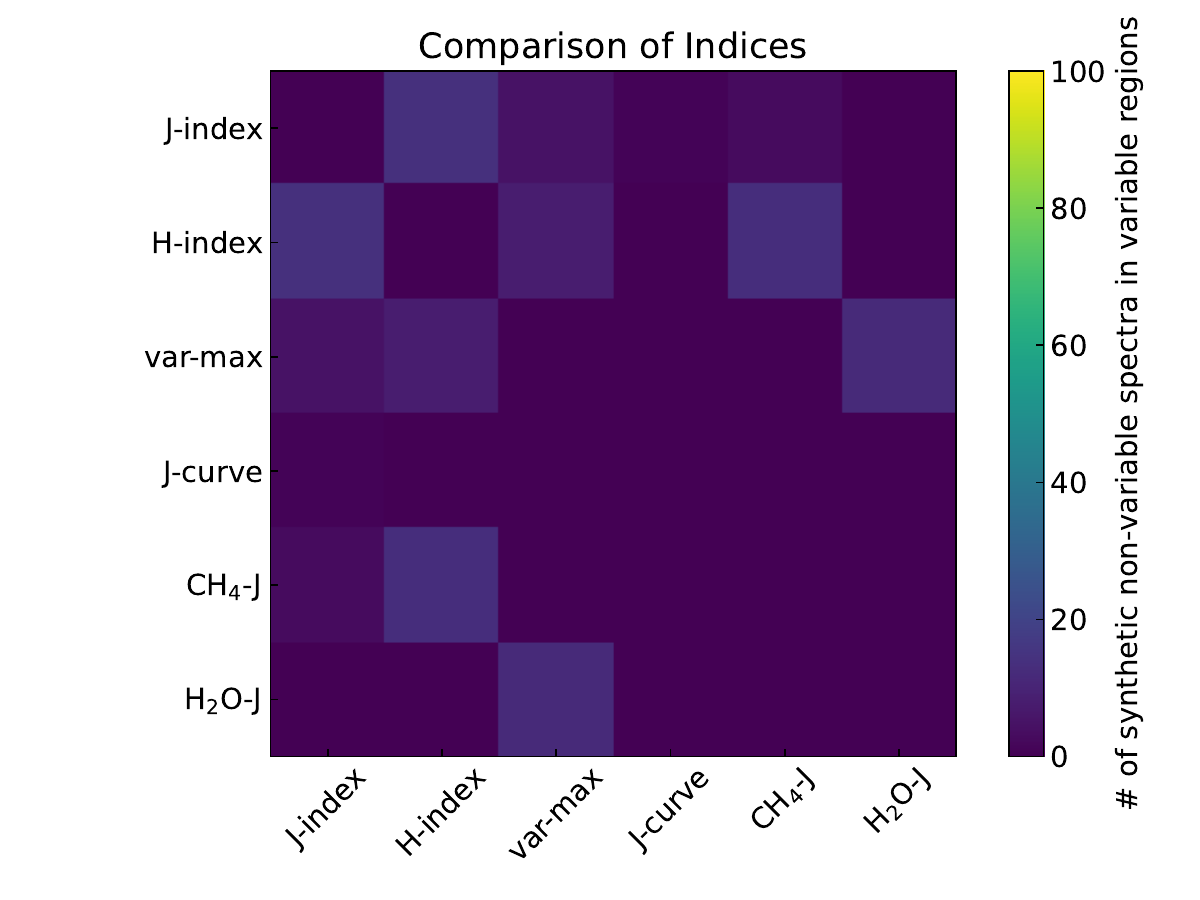}
    \caption{Matrix 6x6 combination of indices between them applied in the synthetic variable (left) and non-variable (right) spectra. The color map represents the number of the 100 synthetic spectra of each sample that fall in variable regions of each combination. In the left plot, we see in yellow the index-index combinations that are best at finding L dwarf variable candidates, and in blue those that are less useful. In contrast, in the right plot, we see that all index-index combinations succeed at not finding any non-variable objects.}
    \label{fig:relevant-synthetic-indices}
\end{figure}

In Figure \ref{fig:relevant-synthetic-indices}, we show the 6x6 matrix of all possible index-index combinations. In this plot, the color map represents the number of variables (Fig. \ref{fig:relevant-synthetic-indices} - left) and non-variable synthetic spectra (Fig. \ref{fig:relevant-synthetic-indices} - right) falling within the variability regions in the index-index plots defined in Section \ref{sec:index-method}.

In Figure \ref{fig:relevant-synthetic-indices}, we noticed that the combination of H$_2$O-J index vs J- and H-indices seem less effective than the others in finding variable brown dwarfs. However,  in Figure \ref{fig:relevant-synthetic-indices}, right we see that all index-index combinations, including H$_2$O-J index vs J- and H-indices can find non-variable spectra. Therefore, we concluded that all index-index combinations of indices are necessary to find either variable or non-variable spectra.

\begin{figure}[h]
    \centering
    \includegraphics[width=0.48\linewidth]{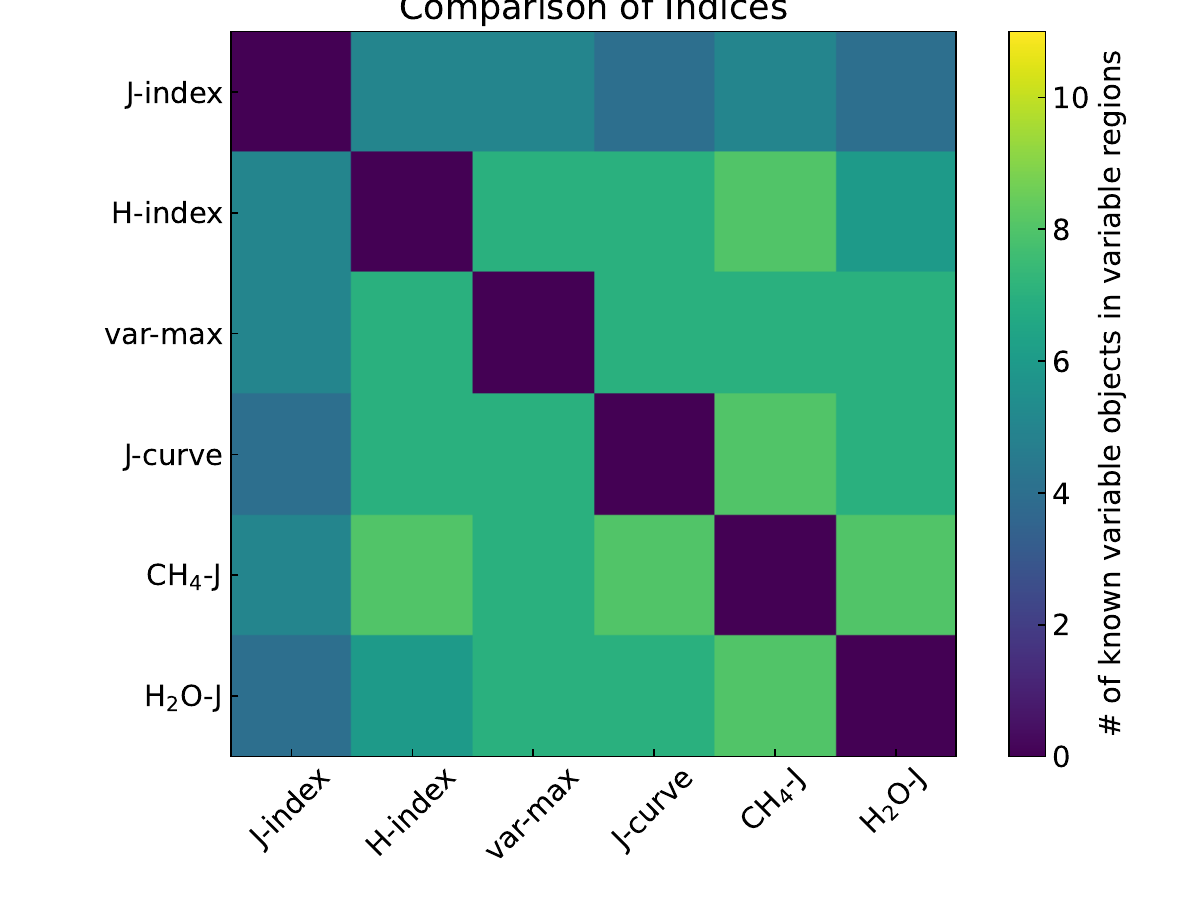}
    \includegraphics[width=0.48\linewidth]{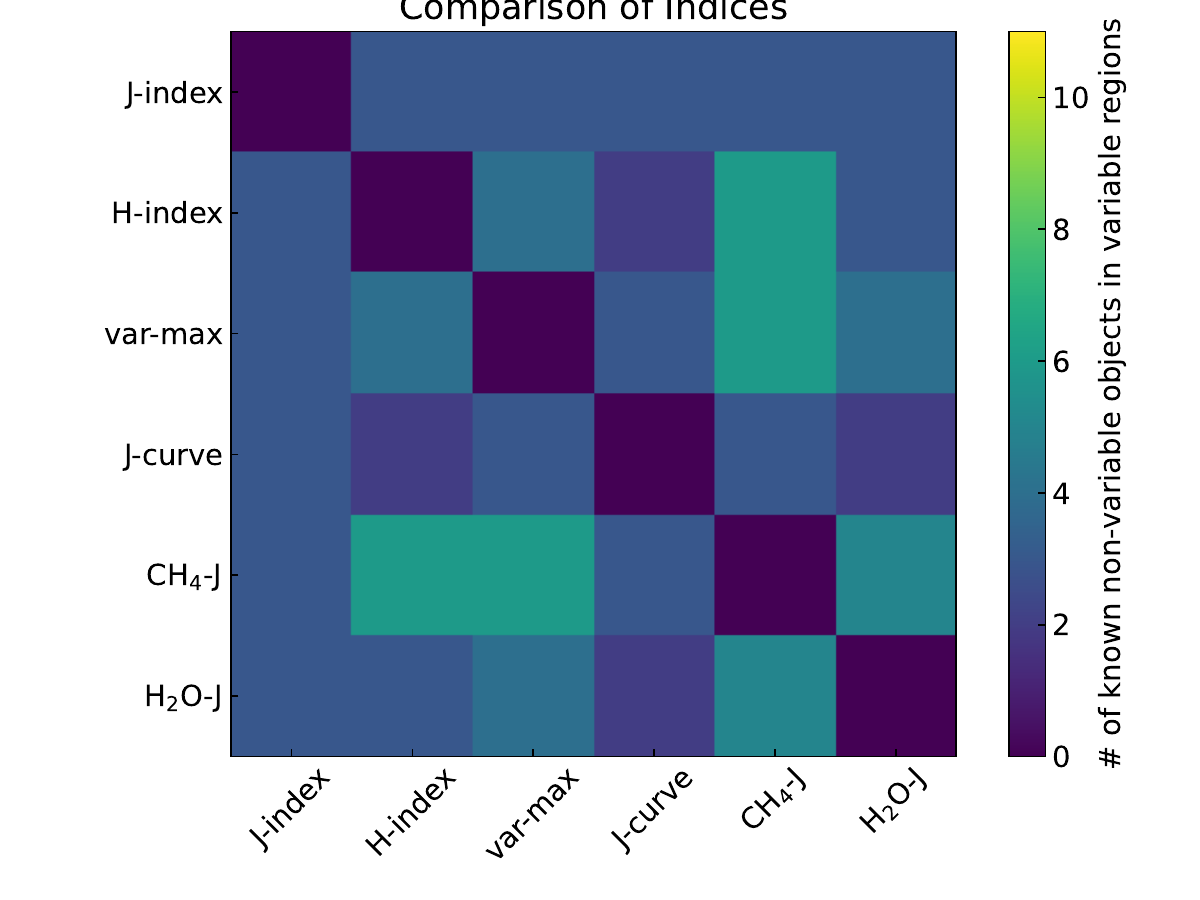}
    \caption{Matrix 6x6 combination of indices between them applied in the 11 known variables (left) and the 9 non-variable (right) spectra in the literature. The color map represents the number of objects of each sample that fall in variable regions of each combination.}
    \label{fig:relevant-knwon-indices}
\end{figure}

Following the same methodology explained previously, we use the spectra of the known variables in the literature (Figure \ref{fig:relevant-knwon-indices}, left), and the spectra of the known non-variable (Fig. \ref{fig:relevant-knwon-indices}, left), and we explore which index-index combinations are able to find most of the known variable and non-variable objects. For this case of known objects, we see that the distribution is a little different since it seems that most index-index combinations are the same efficient at finding variable and non-variable objects.

\end{document}